\documentclass[prx,twocolumn,superscriptaddress,nofootinbib,floatfix]{revtex4-1}
\usepackage{soul}
\usepackage{amsfonts}
\usepackage{graphicx}
\usepackage{bm}
\usepackage{amssymb}
\usepackage{amsmath}
\usepackage{amstext}
\usepackage{latexsym}
\usepackage{enumitem}
\usepackage[usenames,dvipsnames]{xcolor}
\definecolor{crimson}{RGB}{220,20,60}
\newcommand\ba{\begin{eqnarray}}
\newcommand\ea{\end{eqnarray}}
\newcommand\be{\begin{equation}}
\newcommand\ee{\end{equation}}

\newcommand{\ket}[1]{|#1\rangle}

\usepackage{braket}
\usepackage{float}

\usepackage{verbatim}
\usepackage{graphicx}
\begin{document}
\author{Manju C}
\email{222004003@smail.iitpkd.ac.in}
 \affiliation{Department of Physics, Indian Institute of  Technology Palakkad, Palakkad, Kerala, 678623.}
\author{Uma Divakaran}
\email{uma@iitpkd.ac.in}
 \affiliation{Department of Physics, Indian Institute of  Technology Palakkad, Palakkad, Kerala, 678623.}
\author{Arul Lakshminarayan}%
 \email{arul@physics.iitm.ac.in}
\affiliation{%
Department of Physics, \& Center for Quantum Information, Computation and Communication, \\ Indian Institute of Technology Madras, Chennai, India~600036}

%\title{Evolution towards random matrix theory in full Hilbert space by disorder}
\title{Disordering a permutation symmetric system: revivals, thermalization and chaos }

\date{\today}

\begin{abstract}

This study explores the effects of introducing a symmetry breaking disorder on the dynamics of a system invariant under particle permutation. The disorder forces quantum states, confined to the $N+1$ dimensional completely symmetric space  to penetrate the exponentially large $2^N$ dimensional Hilbert space of $N$ particles.
In particular, we focus on the quantum kicked top as a Floquet system of $N$ qubits, and use linear entropy, measuring single qubit entanglement, to investigate the changes in the time scales and values of saturation when disorder is introduced. In the near-integrable regime of the kicked top, we study the robustness of quantum revivals to disorder. We also find that a classical calculation yields the quantum single qubit entanglement to remarkable accuracy in the disorder free limit.  The disorder, on the other hand, is modeled in the form of noise which again fits well with the numerical calculations. We measure the extent to which the dynamics is retained within the symmetric subspace and its spreading to the full Hilbert space using different quantities. 
We show that increasing disorder drives the system to a chaotic phase in full Hilbert space, as also supported by the spectral statistics.
We find that there is robustness to disorder in the system, and this is a function of how chaotic the kicked top is.

\end{abstract}


\maketitle
%%%%%%%%%%%%%%%%%%%%%%%%%%%%%%%%%%%%%%%%%%%%%%%%%%%%%%%%%%%%%%%%%%%%%

\section{Introduction}
Symmetry of a Hamiltonian plays a crucial role in the stationary states and dynamics of the system. It is related to various conserved quantities which also helps in making calculations simpler. In this work, we focus on breaking the permutation symmetry of a system whose Hamiltonian is
invariant under the exchange of its entities. This symmetry can be broken 
say by introducing disorder, as done in this work. 
The disorder free limit can show regular or chaotic dynamics depending on the nonlinearity of the system. Thus the permutation symmetry breaking disorder's  interplay with the deterministic chaos of the system is of interest. Specifically, we explore the possibility of regular dynamics  morphing to quantum chaos as the disorder strength is increased. 

Understanding the emergence of quantum chaos is naturally important owing to its effect on practical implementation of various condensed matter systems. For example, chaos can induce decoherence 
%in a system, thus questioning 
and the reliability of the system in various quantum information processing applications \cite{zurek1995quantum, chaos_border} can be doubtful. Several other aspects connecting chaos with quantum computing have been studied in Refs. \cite{miller_coupled_top, bandyopadhyay2002testing, bandyopadhyay2004entanglement, lakshminarayan2001entangling, Decoherence_dicke_model, flambaum2000time, song2001quantum, georgeot2000emergence, braun2002quantum, madhok2018quantum, chaos_and_computers, piga2019quantum}, for example it has been shown in Ref. \cite{LMG_PRR} that the presence of chaos can lead to proliferation of errors restricting the ability to get correct output from noisy intermediate scale quantum (NISQ) devices.

In classical chaos, the phase-space trajectories and their Lyapunov exponent are used as an indicator of chaos \cite{Haakebook, Sandrobook}. %Extending chaos to the quantum world is a non-trivial task since the unitary evolution can not capture the sensitivity to the initial conditions, the way it is defined in classical chaos.
For quantum chaos, alternate methods include studying energy level statistics that shows Poisson distribution when the dynamics is regular and Wigner Dyson distribution when chaotic \cite{Haakebook, spacing_statistics, wang2023statistics}. Other measures include the temporal evolution of various expectation values \cite{Haakebook, otocandloschmidt, papparaldi_bridging}, entanglement measures such as von Neumann entropy, concurrence, discord and so on \cite{tripartitescrambling, concurrence, wang2002pairwise, chaudhury_nature_husimi}.
At the same time, various other works have also explored how the existence of disorder plays a crucial role in safeguarding individual qubit states from non-linear effects in a non-linear quantum resonator \cite{Berke_2022, taming_disorder,trust_quantum_simulators}.
Therefore, it is interesting and important to study the effect of disorder on the dynamics of a system that could be regular or chaotic. 

We use the well-studied quantum kicked top, which is a periodically forced infinite range interacting system consisting of $N$ qubits 
which has a permutation symmetry under the exchange of particles, where the dynamics is restricted in the $N+1$ dimensional Hilbert space as opposed to the full Hilbert space of $2^N-$dimensions \cite{baguette2014multiqubit, wang2002pairwise, devi2012majorana, markham2011entanglement, popkov2012reduced, LMG_intro, russomanno2015thermalization, tripartitescrambling, russomanno2021quantum, papparaldi_bridging, vidal2004_TSS}. 
The previous studies on quantum kicked top model consist of finding relation between chaotic dynamics and entanglement measures \cite{concurrence, decoherence, pattnayak}, studying periodicity of the Floquet operator of the top \cite{periodicity}, signatures of bifurcation on quantum entanglement measures \cite{bifurcation_QKT} and employing random matrix theory (RMT) to study the chaotic limit of the quantum top \cite{tripartitescrambling, generalized_ent}. 
All these studies are in the symmetric basis, also called the Dicke basis, that naturally arises in this multi-qubit system \cite{tripartitescrambling, wang2002pairwise}. 
Experimental realizations of quantum kicked top as a multi qubit system has been carried out using laser-cooled Cs atoms by utilising their combined electronic and nuclear spin \cite{chaudhury_nature_husimi}, by making use of the interaction of three superconducting qubits in a coupling circuit \cite{2016_ergodic}, by  using nuclear magnetic resonance techniques to mimic two qubit system comprising two spin 1/2 nuclei \cite{NMRstudiesin2qubit}.
The kicked top has been recently used to 
perform quantum metrology and the use of quantum chaos has promoted the approach to the Heisenberg limit that is used for quantum sensing \cite{quantummetrology}. It is therefore of interest in this context as well to study the effects of the breaking of permutation symmetry by disorder. \textcolor{black}{Several studies have investigated the impact of disorder on breaking of this symmetry, for example see Ref.~\cite{random_spin_models, Dicke_auditya}.}

In a recent work \cite{manju24}, we showed the possible presence of a continuous phase transition, using the total angular momentum as the order parameter, from a dynamics restricted within symmetric subspace to a dynamics extending to the full Hilbert space of $2^N$ dimensions as the disorder strength is increased.
In the present paper we study primarily the single qubit entanglement via the linear entropy and the changes in the time scales involved with chaos and disorder. Two additional measures of symmetry breaking are also studied via the overlap of time evolving states with the symmetric subspace as well as an effective dimension involving the eigenstates of Floquet operator. Their convergence to the random matrix limit in full Hilbert space is examined.

A more detailed outline of the paper is as follows. Section \ref{sec_model} presents a broad overview of the kicked top model. This is followed by a detailed analysis in Sec.~\ref{pi/2} of the classical dynamics as well as the timescales associated with growth and saturation of single qubit linear entropy in the quantum regime. Sec.~\ref{sec_disorder_interaction} deals with the main part of the work, which involves breaking the permutation symmetry in the model by introducing disorder, as a result of which the dynamics span the entire Hilbert space of $2^{N}$ dimensions. We study various quantities to understand this change in dynamics like the overlap $\chi$, effective dimension and the long time averaged linear entropy. In Sec.~\ref{4pi/11}, we tweak a particular parameter minimally that enables the study of traditional spectral statistics, and also note that the breaking of permutation symmetry leads to unusual density of state that needs further study. We briefly discuss the effect of disorder in the field term of the Hamiltonian using long time linear entropy in Sec.~\ref{sec_disorder_field} and we summarize and discuss future directions in Sec. \ref{conclusion}.

\section{The kicked top models}
\label{sec_model}
The infinite range Hamiltonian of $N$ spin-1/2 particles in presence of periodic kicks is given by:
\begin{equation}
H=\frac{k}{\tau4N }\sum_{i,j=1}^{N}\sigma_{i}^{x}\sigma_{j}^{x}  +\frac{p}{2}\sum_{n=-\infty }^{\infty }\sum_{i=1}^{N}\sigma_{i}^{y}\delta (t-n\tau),
\label{eq_spin}
\end{equation}
where $\sigma_i^{x}$, $\sigma_i^{y}$ are the Pauli operators acting at site $i$ that obeys the commutation relation $[\sigma^{p},\sigma^{q}]=2i\epsilon_{pqr}\sigma^{r} $.
The first term denotes all-to-all two body interactions between spins in $x$ direction, followed by a periodically kicked field in $y$ direction. Here, $k$ is the chaos parameter that is tuned to undergo a transition from regular ($k=0$) to chaotic dynamics ($k\gg 3$) in the classical or mean-field limit, defined later and $\tau$ is the time period between the kicks. 
The above Hamiltonian can be written in terms of collective angular momentum operators by
defining  $J_{\alpha}$=$\sum_{i=1}^{N}\sigma_{i}^{\alpha}/2$ \cite{milburn1999simulating, concurrence}, where $\alpha=x,y,z$, so that
%and $N=2j$ where j is the total angular momentum quantum number.
%The Hamiltonian also called as the kicked top model is given by:
\begin{equation}
H=\frac{k}{N\tau }J_{x}^2 + p J_{y}\sum_{n=-\infty }^{\infty }\delta (t-n\tau),
\label{eq_kicked}
\end{equation}
also known as quantum kicked top \cite{haake1987classical, concurrence, tripartitescrambling}.

The angular momentum operators obey the commutation relation $[J_{i},J_{j}]=i\hbar \epsilon_{ijk}J_{k} $. Since $[J^{2},H]=0$, $\langle J^{2} \rangle =j(j+1)\hbar^{2}$ is a conserved quantity where $j$ is the angular momentum quantum number. Both the classical limit ($j \to \infty$) and quantum limit (finite $j$) of kicked top has been extensively studied in the context of chaos and quantum information measures \cite{haake1987classical, effect_measurement_qkt, concurrence, spinsqueezing, decoherence, periodicity, periodic_orbits, fewbodykickedtop, bifurcation_QKT, madhok_quantum_discord, chaudhury_nature_husimi, zou2022pseudoclassical, lombardi2011entanglement}. If we initialize the system in a spin coherent state \cite{Haakebook, spincoherentstates, glauber_spin_coherent_state} given by
\begin{equation}
\ket{\psi_0}=\ket{\theta,\phi}=  \left (\cos\frac{\theta}{2}\ket{0}+e^{i\phi}\sin\frac{\theta}{2}\ket{1}\right)^{\otimes{N}},
\label{eq_coherent}
\end{equation}
which corresponds to $j=N/2$, the dynamics is restricted to $2j+1$ (or $N+1$) dimensional Hilbert space due to conservation of $J^2$. 
This is the state with minimum uncertainty and hence closest to the classical kicked top  defined by $(j \sin \theta \cos \phi, j \sin \theta \sin \phi, j \cos \theta)$ \cite{Haakebook, decoherence}.
Such an initial state helps in comparing dynamics of the quantum system with the results of the corresponding classical system, which is one of the goals of the current work.
The evolution of the kicked top or kicked spin chain can be studied using Floquet operator $U_0$ which determines the evolution of the state from one kick to the time just before the next kick, and is given by
\begin{equation}
U_0= \exp\left({-i\frac{k}{2j}J_{x}^{2}} \right) \hspace{0.1 cm} \exp\left({-i pJ_{y}}\right),
\label{eq_floquet_cl}
\end{equation}
where we have put $N=2j$.  We have also set $\tau=1$ and $\hbar=1$  so that the effective Planck's constant is given by $1/j$.

The primary focus of this paper is to understand how the presence of disorder in the interaction term affects the regular or chaotic nature of the all-to-all spin Hamiltonian (Eq.~\ref{eq_spin}). The disordered version of Eq.~\ref{eq_spin} is given by
\begin{equation}
H=\frac{k}{\tau4N }\sum_{i,j=1}^{N}(1+\epsilon_{ij})\sigma_{i}^{x}\sigma_{j}^{x}  +\frac{p}{2}\sum_{n=-\infty }^{\infty }\sum_{i=1}^{N}\sigma_{i}^{y}\delta (t-n\tau),
\label{eq_spin_disorder}
\end{equation}
where $\epsilon_{ij}$ is a random number taken from a normal distribution with zero mean and standard deviation $w$. We denote the disorder free case as $w=0$ and disordered case as $w\ne 0$.  Clearly, in the disordered case, $J^2$ is no longer a constant of motion and the dynamics will explore the full Hilbert space.
We first discuss the well known results from classical phase space (meaningful {\it only} in the $w=0$ limit) analysis of the kicked top \cite{Haakebook, haake1987classical} followed by the results already known in the corresponding spin system when studied using quantum measures \cite{concurrence,decoherence}.  Later we compare the effects of disorder on these quantum measures using Eq.~\ref{eq_spin_disorder}. We first present the results of classical and quantum dynamics for the special case of $p=\pi/2$, both in the presence and in the absence of the disorder in section \ref{pi/2}. In section \ref{4pi/11}, we show that the results are robust for other values of $p$ as well, taking the example of $p=4\pi/11$.

\section{Disorder free, $w=0$, and  $p=\pi/2$}
\label{pi/2}
\subsection{Classical dynamics}
\begin{figure}[h]
\includegraphics[scale=0.2]{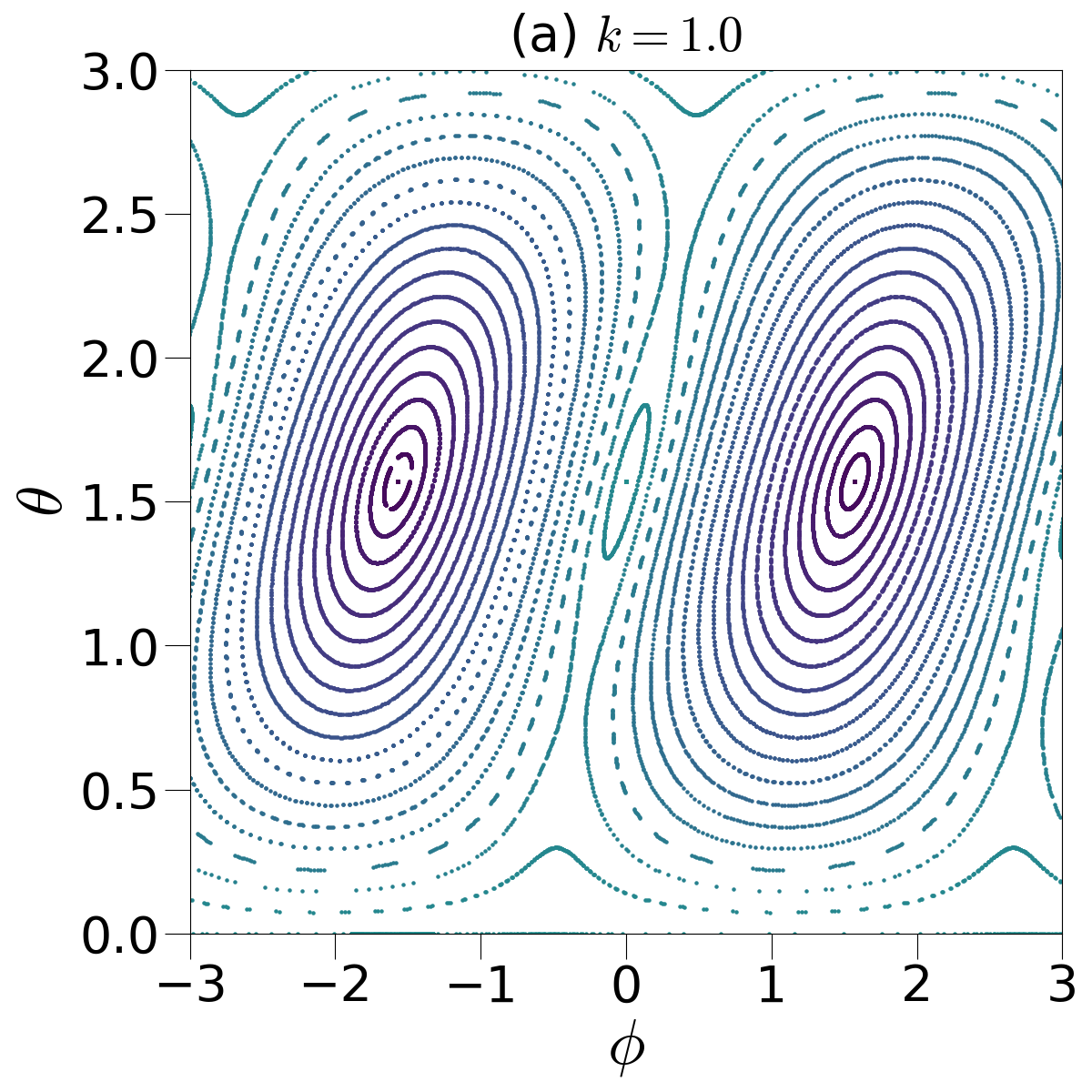}
\includegraphics[scale=0.2]{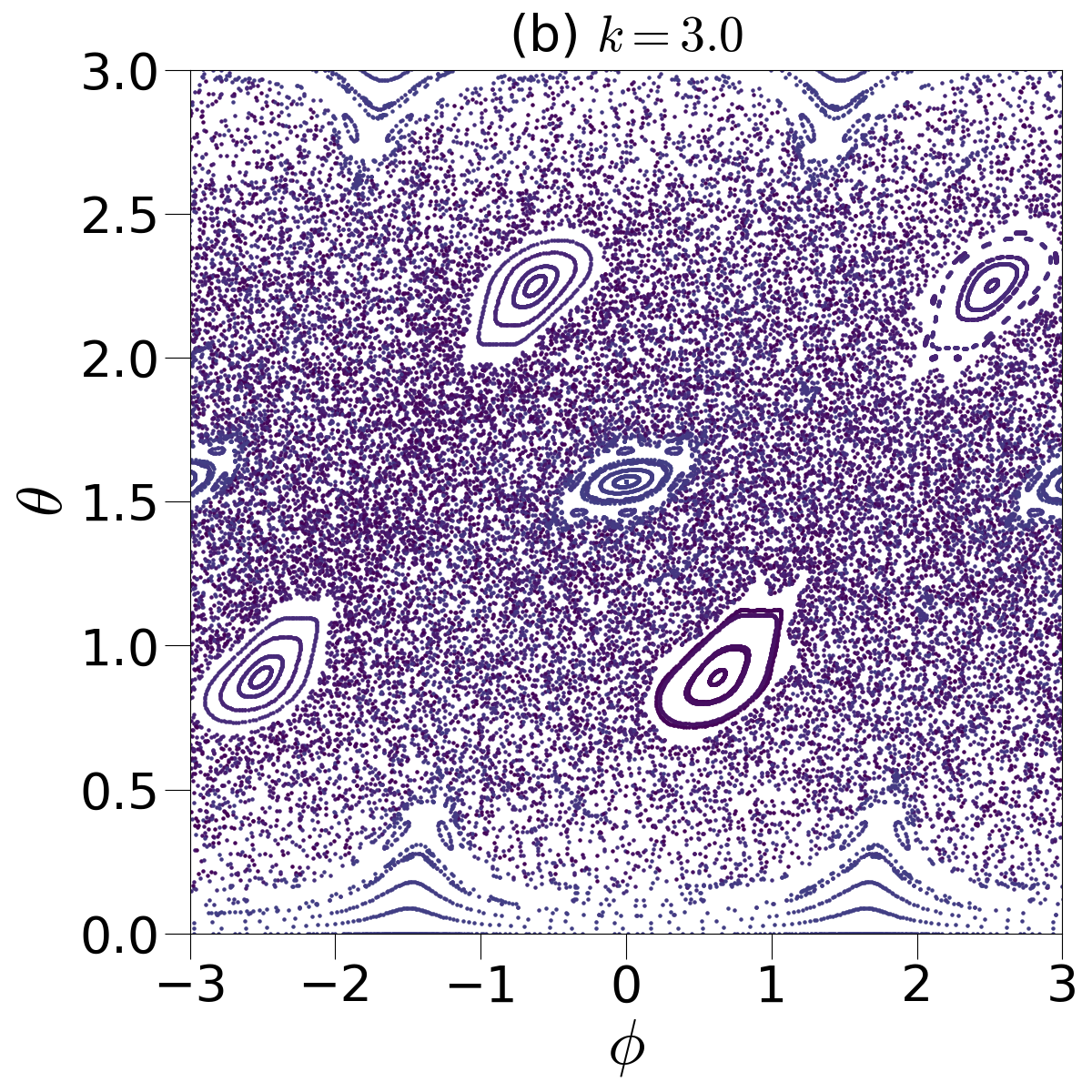}
\includegraphics[scale=0.2]{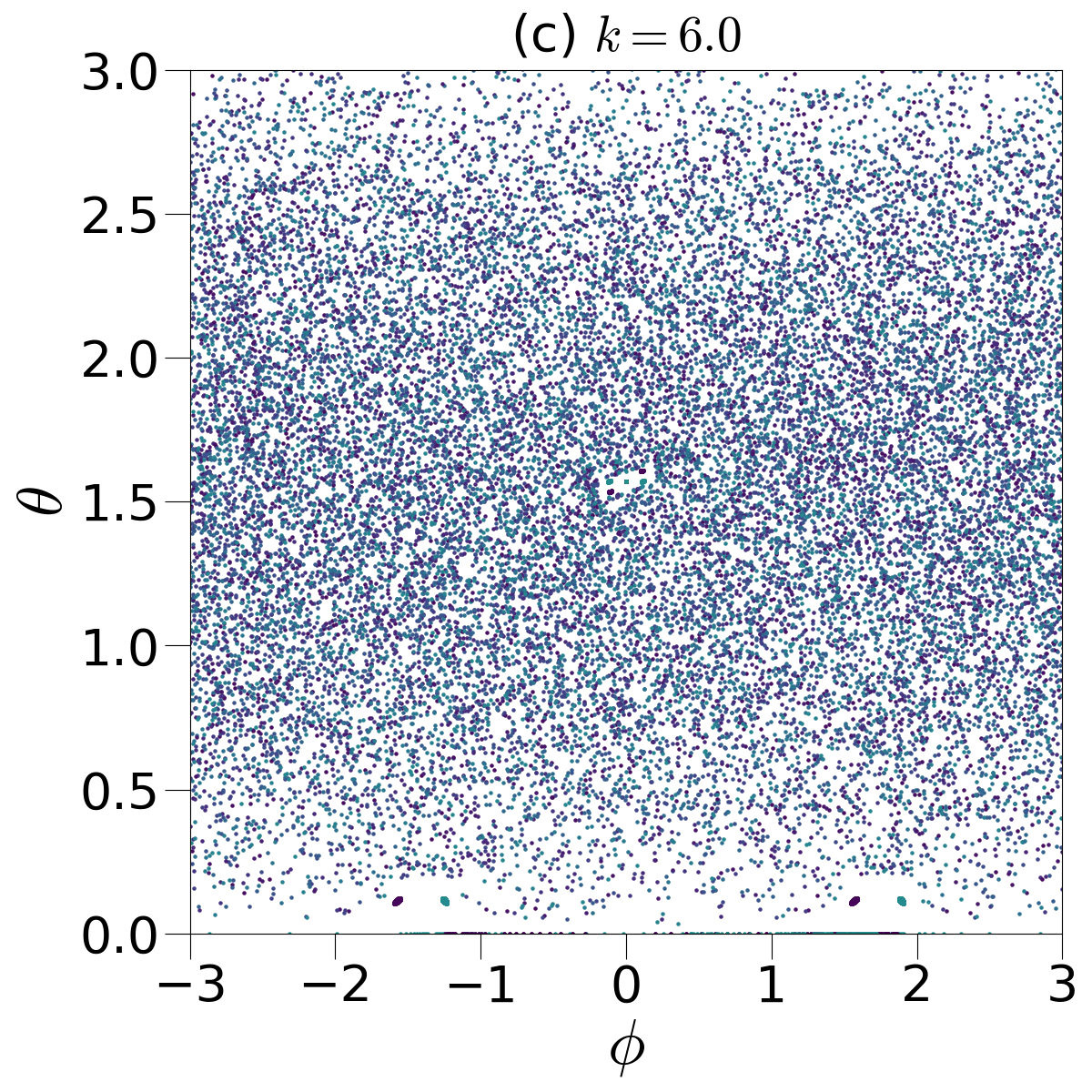}
\caption{The classical phase space of kicked top model for different chaos parameter $k$. (a) $k=1$ shows predominantly regular islands, (b) $k=3$ corresponds to mixed phase space and (c) $k=6$ is fully chaotic. 
}
\label{classical_phase_space}
\end{figure}

The Heisenberg evolution equation connecting expectation values at $n$-th kick and $n+1$-th kick is given by \cite{decoherence}
\begin{equation}
 (J_{i})_{n+1}=U_0^{\dagger} \left(J_{i} \right)_{n} U_0,
\end{equation}
where $U_0$ is the evolution operator as given in Eq.~\ref{eq_floquet_cl}. The classical limit is obtained by taking the limit $j\rightarrow \infty$  which corresponds to  $N\rightarrow \infty$ in the spin language. 
We rescale the angular momenta as $X=\frac{J_{x}}{j}$, $Y=\frac{J_{y}}{j}$, $Z=\frac{J_{z}}{j}$ and  obtain the following classical equations of motion  in the  limit $j\rightarrow \infty$, :
\begin{eqnarray}
X_{n+1}&=&Z_{n} \nonumber ,\\
Y_{n+1}&=&Y_{n}\cos(kZ_{n})+X_{n}\sin(kZ_{n}) \nonumber ,\\
Z_{n+1}&=&-X_{n}\cos(kZ_{n})+Y_{n}\sin(kZ_{n}) \nonumber .\\
\label{map}
\end{eqnarray}
Fig.~\ref{classical_phase_space} shows the classical phase space of the kicked top model for different ``chaos parameter" $k$. Here $\theta = \cos^{-1}(Z)$ and $\phi = \tan^{-1}(\frac{Y}{X})$.  As we increase the chaos parameter $k$ of the Hamiltonian, the system dynamics changes from regular to chaotic behavior.

The salient features of the phase space structures are:
the phase space for $k=1$ has predominantly regular region with periodic orbits; for $k=3$, it shows mixed phase space with regular islands in a sea of chaotic trajectories, whereas $k=6$ is fully chaotic.
In order to analyze these phase space plots further, we write Eq.~\ref{map} as $[X_{n+1},Y_{n+1},Z_{n+1}]= F [X_{n},Y_{n},Z_{n}]$, where $F$ is a map that takes the variable from $n$-th time step to $n+1$-th time step. Fixed points are defined as the points when $F(X,Y,Z)=(X,Y,Z)$ which gives ($\frac{\pi}{2}$,$\frac{\pi}{2}$) and ($\frac{\pi}{2}$,-$\frac{\pi}{2}$) as the fixed points for $k=1$, as also clearly depicted in Fig.~\ref{classical_phase_space}(a). The solutions to the equation $F^{n} [X,Y,Z]= [X,Y,Z]$ gives the period $n$ orbits of the system \cite{haake1987classical, periodic_orbits, kickedpspin}. When the chaos parameter is increased to $k=3$, the fixed points ($\frac{\pi}{2}$,-$\frac{\pi}{2}$) and ($\frac{\pi}{2}$,$\frac{\pi}{2}$) bifurcate giving rise to period 2 and two period 1 orbits respectively as shown in Fig.~\ref{classical_phase_space}(b). %We can also see the presence of period 4 orbits at ($\frac{\pi}{2}$,$\pm \pi $) and ($\frac{\pi}{2}$,0) in Fig.\ref{classical_phase_space}.(b). 
There also exists period 4 orbits at ($\frac{\pi}{2}$,$\pm \pi $), ($\frac{\pi}{2}$,0), ($\pi$,$\phi$) and (0,$\phi$) which are stable till $k=\pi$. These points or orbits loses stability as the chaos parameter $k$ is further increased.  
\subsection{Quantum dynamics}
\label{quantum_dynamics_0var}
We now discuss the quantum kicked top, i.e., kicked top with finite $N$ using quantum measures. The state $|\psi_n\rangle$ after $n$ kicks starting from a spin coherent state $|\theta,\phi\rangle$ is given by $|\psi_n\rangle= U_0^n |\theta,\phi\rangle$. As mentioned before, the dynamics of this spin  system has been studied using linear entropy, concurrence, tripartite mutual information, out of time ordered correlator measures \cite{concurrence, otocandloschmidt, tripartitescrambling, madhok_quantum_discord, spinsqueezing, periodicity, chaudhury_nature_husimi, quantummetrology}. Here, we focus on one such measure, namely, linear entropy especially focusing on its long time averaged behavior and short time behavior. Linear entropy $S_Q$ between a bipartition of $Q:N-Q$ qubits for an $N$ qubit system, is given by:
\begin{equation}
S_{Q}= 1- \text{Tr}\,\rho_{Q}^2,
\label{linear_entropy}
\end{equation}
where $\rho_{Q}$ is the reduced density matrix of $Q$ qubits. Below, we discuss various time scales in the disorder free problem using linear entropy, and later compare them when disorder is introduced.

\subsubsection{Temporal behavior of linear entropy:\\ Ehrenfest and Revival time scales}
Ehrenfest time scale \cite{berry_balaz_1979, papparaldi_bridging, ehrenfest_algebraic, Alessio_ehrenfest} denoted as $\tau_{\hbar}$ is defined as the time scale over which the classical and quantum dynamics of a system remains comparable, i.e., till the time when quantum interference effects start manifesting.
In the near-integrable regime when $k \ll 3$, apart from the Ehrenfest time scale, wavepacket revivals can occur over a much longer time scale called as the revival time $t_{r}$. This revival occurs when the interference is constructive enough to result in fractional or full revivals of wavepackets \cite{robinett2004quantum, Fishman_revival, zhao2019quantum}. In general, revivals are not observed in chaotic regime \cite{tomsovic1997can, Ent_revivals}. Ehrenfest time $\tau_h$ can be estimated as the time at which the quantum fluctuations become comparable to the typical length $(1/\hbar)$ in phase space. It is well established that for integrable systems where quantum fluctuations grow polynomially in time $\sim t^2$, $\tau_{\hbar} \sim 1/\sqrt{\hbar}$ (or some powers of $\hbar$) and for the chaotic case where there is exponential growth $\tau_{\hbar} \sim \ln(1/\hbar)$ \cite{papparaldi_bridging, berry_balaz_1979}.
 
 For the case of kicked top $1/{\hbar} \sim N$ so that $\tau_{\hbar}\sim \sqrt{N}$ in the regular region and $\tau_{\hbar}\sim \ln {N}$ in the chaotic region. We use linear entropy  to extract the Ehrenfest time.
 The single qubit linear entropy can be written in terms of the collective angular momentum operators as \cite{decoherence}:
 \begin{eqnarray}
 S_1(n)=\frac{1}{2}\left[1-\left(\frac{\langle J_x(n) \rangle^2 +\langle J_y(n) \rangle^2+ \langle J_z(n) \rangle^2}{j^2}\right)\right]
 \label{entropy_eq}
 \end{eqnarray}
 where the expectation values are evaluated for a given initial state.

 It is straightforward to find the classical limit of this. Noting that $X_n^2+Y_n^2+Z_n^2=1$, the classical unit sphere of states, we can write the corresponding classical linear entropy as 
 \begin{equation}
 S_{cl}(n) =\frac{1}{2} \left( \langle \Delta X_n^2\rangle + \langle \Delta Y_n^2\rangle+\langle \Delta Z_n^2\rangle \right)
 \label{eq:ClassicalEnt}
 \end{equation}
 Here $\langle \Delta X_n^2\rangle$ is the variance of $X_n$ over an ensemble of initial states, and so also for $Y_n$ and $Z_n$. Thus the classical linear entropy vanishes when the ensemble becomes a ``pure state" of a delta function, and is nonzero when the ensemble is spread. In other words, it measures how much the ensemble is delocalized. \textcolor{black}{We note the somewhat paradoxical feature that there is a well-defined classical limit for the entanglement of a single qubit. This is due to the permutation symmetry of the system wherein the collective spin expectation value is $N$ times that of any one spins'.}
 
 \begin{figure}
\includegraphics[width=0.9\linewidth, height=0.8\linewidth]{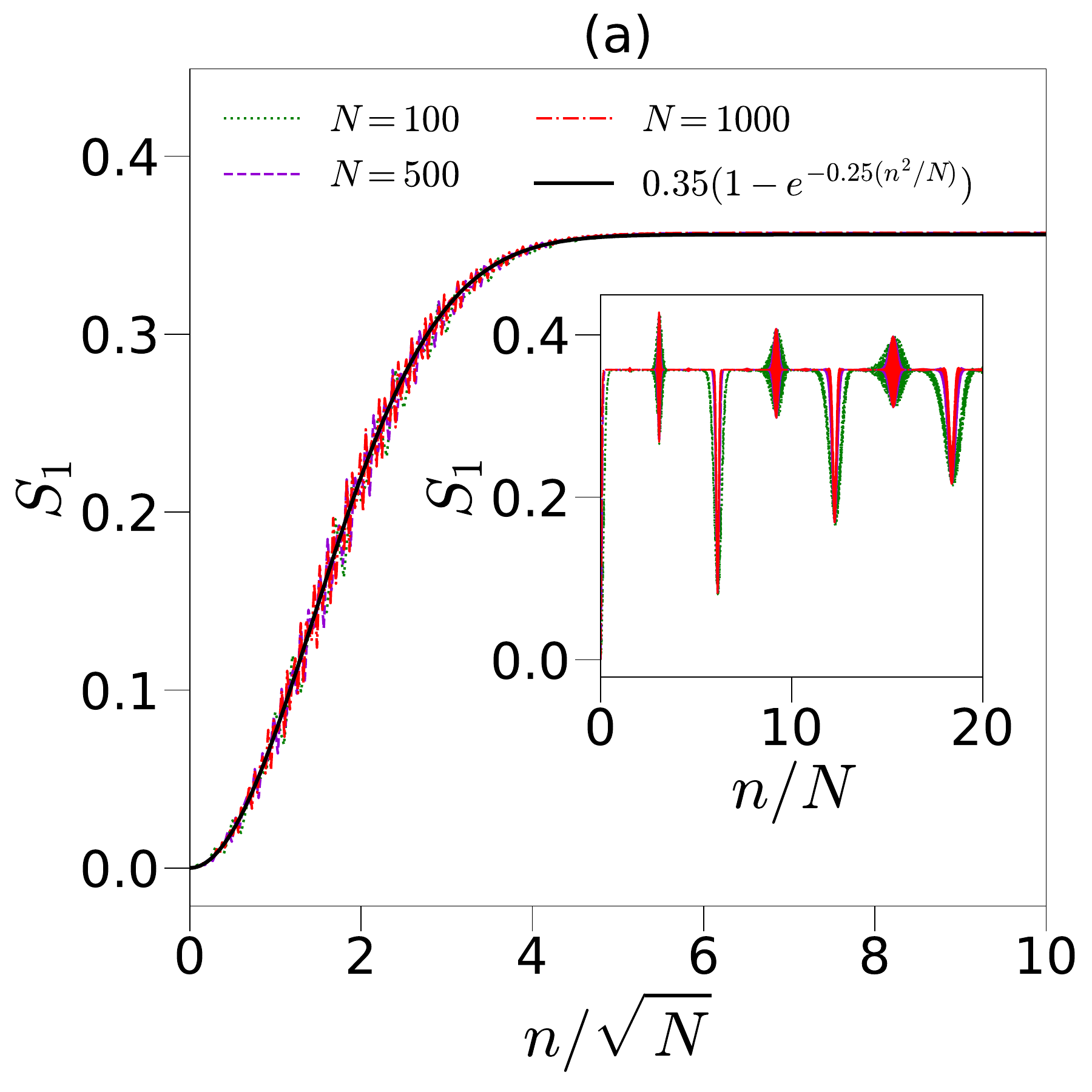}
\includegraphics[width=1.0\linewidth, height=0.8\linewidth]{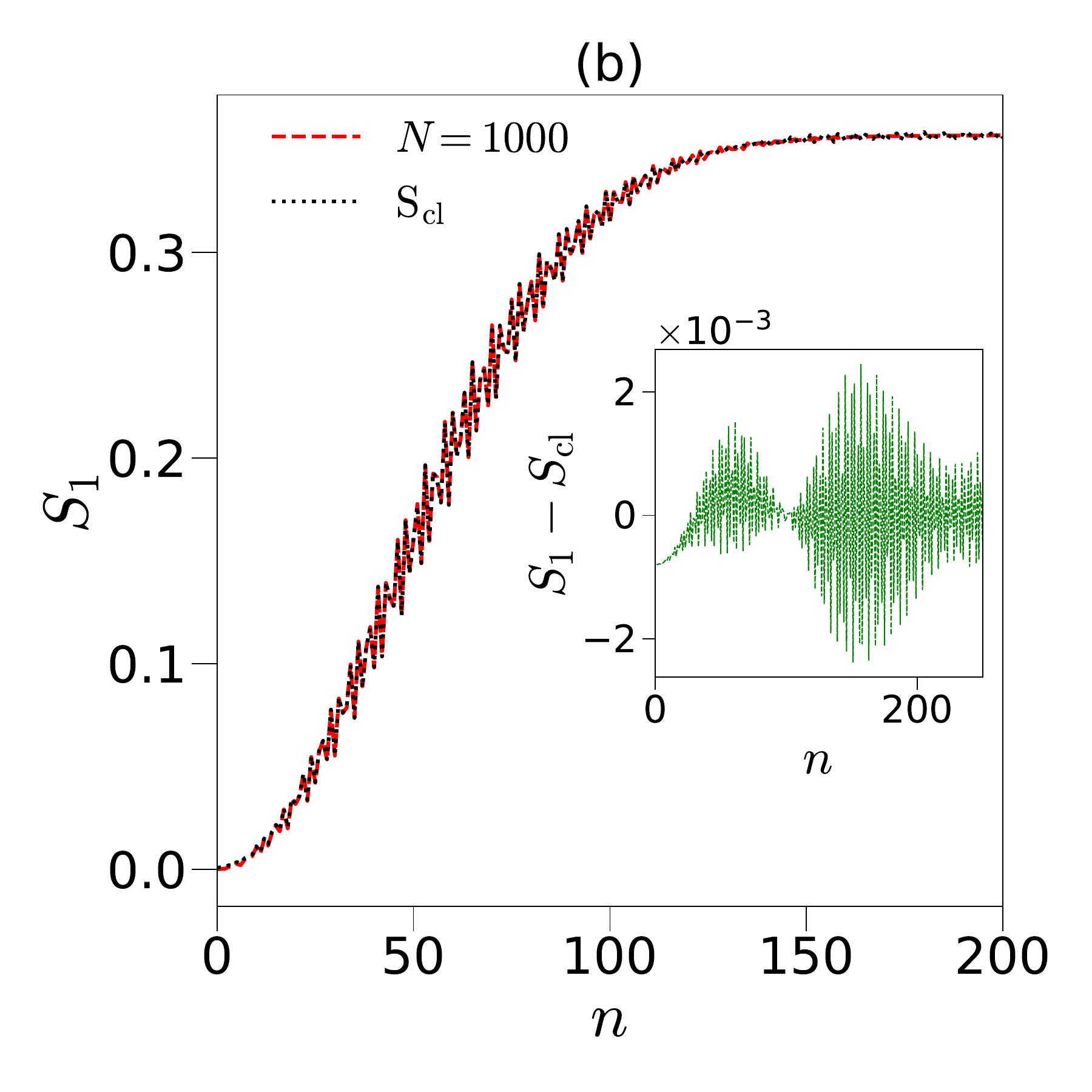}
\caption{(a) $S_1$ as a function of $n/\sqrt{N}$ for different $N$ at $k=1$ for the state $\ket{2.25,1.1}$. The collapse of different figures at initial short times confirm $\tau_{\hbar}\sim \sqrt{N}$. Inset shows the scaling of $t_r\sim N$ where we observe collapse of revivals when $S_1$ is plotted as a function of $n/N$ at long times. (b) The classical and quantum correspondence for $k=1$ with initial condition $(2.25,1.1)$. The red dashed line is obtained using quantum dynamics and the black dotted line is obtained using classical ensemble as discussed in the text. The inset highlights the small difference (of the order of $10^{-3}$) between single qubit linear entropy in the quantum limit and using classical ensemble. }

  \label{revival_time_0var}
\end{figure}

 Since the classical entropy for a pure state is zero, the growth of $S_1$ for finite $N$ quantum system essentially captures the Ehrenfest time. In the regular region of small $k$ values, we obtain a good fit of the numerical data  with the function
 \begin{equation}
 S_1(n)=S_{\infty}\left (1-\exp\left(- \alpha\,  n^2/N \right)\right).
 \label{eq:regular_fit}
 \end{equation}
Thus, $S_1$ is zero for $n \ll \sqrt{N}$, or comparable to classical value, whereas it saturates to a nonzero value for $n \gg \sqrt{N}$, confirming $\tau_{\hbar} \sim {\sqrt{N}}$.
 Fig.~\ref{revival_time_0var}(a) shows the collapse of $S_1$ for different $N$ when the time step is scaled as $n/\sqrt{N}$ highlighting $\tau_{\hbar} \sim \sqrt{N}$. It also plots the fitted expression in Eq.~\ref{eq:regular_fit} with $\alpha \approx 0.25$ and the saturation value $S_{\infty} \approx 0.35$ that perfectly matches the numerical data.
 Here the initial state is taken to be 
 $|\theta_0,\phi_0 \rangle= |2.25,1.1 \rangle$. The revival time 
 $t_r$ on the other hand, for integrable systems scales as $t_r \sim \hbar^{-1}$ \cite{robinett2004quantum, Fishman_revival, zhao2019quantum}. Thus in our case, the revivals is expected to show $t_r \sim N$ for larger $n$, which is captured by the collapse of revivals when plotted with $n/N$, as shown in the inset of Fig.~\ref{revival_time_0var}(a).

 We use the other limit of Eq.~\ref{eq:ClassicalEnt} where the ensemble is not localized at a point to study quantum classical correspondence. For this, we evaluate Eq.~\ref{eq:ClassicalEnt} by considering an ensemble of points taken from a Gaussian distribution centered at $(\theta_0,\phi_0)=(2.25,1.1)$ with a standard deviation of $\textcolor{black}{\sigma}=1/\sqrt{2j}$ \cite{fox1994chaos} and evolve these using classical equations of motion given in Eq.~\ref{map}. Compared to distributions like the box-type or circular disk, the Gaussian distribution exhibits excellent quantum-classical correspondence, as also indicated in \cite{fox1994chaos}. The comparison of linear entropy calculated using quantum dynamics and linear entropy $S_{cl}$ calculated using a classical ensemble centered at the same initial state as that of quantum evolution is shown in Fig.~\ref{revival_time_0var}(b). {We see that this correspondence exists even for larger times beyond the Ehrenfest time scale.} The inset shows the difference between the two which is of the order of $10^{-3}$. This remarkable agreement between a purely classical quantity and the quantum entanglement, prompts a closer analysis.

Stepping back from the specific example of the kicked top, consider the classical dynamics of a near-integrable system, which could be a one-degree of freedom flow or a two-dimensional map. The near-integrable regime implies the presence of good, or approximately good, action-angle variables $(I,\theta)$, such that $I$ is a constant, while $\theta(t)=\theta(0)+\omega(I) t$. Let $\rho(\theta,I,0)$ be the initial ensemble, which evolves to $\rho(\theta,I,t)=\rho(\theta-\omega(I)t, I,0)$ according to Liouville's theorem. For the Gaussian density we take
\begin{equation}
    \rho(\theta,I,0)=\frac{1}{2 \pi \sigma^2 } \sum_{l=-\infty}^{\infty}e^{-(\theta +2 \pi l)^2/(2 \sigma^2)}e^{-(I-I_0)^2/(2 \sigma^2)},
\end{equation}
where for simplicity we have taken the same width for both the Gaussians and that the center of the angle variable has been chosen as $0$, with $-\pi \leq \theta <\pi$. The actions are centered tightly around $I_0$. Images of the densities in the angle variable is chosen so that the density is strictly periodic $\rho(\theta,I,0)=\rho(\theta+2 \pi ,I,0)$, however the normalization is calculated as if there is a single Gaussian in the infinite line, as we assume $\sigma \ll 2 \pi$. Given any observable $f(\theta)$, we seek to find the variance 
\begin{equation}
    \text{var}f(t)=\langle f^2\rangle(t)-\langle f \rangle^2(t),
\end{equation}
 where
 \[ \langle f \rangle(t) = \frac{1}{2 \pi}
 \int_{-\pi}^{\pi} \int_{-\infty}^{\infty} f(\theta) \rho(\theta,I,t)\, d \theta\, dI. \]
 Note that this quantity will start out being small and $\rightarrow 0$ as $\sigma \rightarrow 0$, but as time progresses and the density gets smeared over a small neighborhood of invariant tori, it will ``equilibrate" and we wish to track this evolution. Related observations in the context of the kicked top have been made earlier in \cite{generalized_ent}. We will use classical $\text{var} f(t)$ expressions to fit the quantum single qubit entanglement, via the linear entropy.
 
 The equilibration is expected in all nonlinear systems wherein $\omega(I)$ is not a constant, or $\omega'(I_0) \neq 0$ (and hence not a harmonic oscillator or such isochronous systems). Taylor expanding 
 $\omega(I)=\omega(I_0)+\omega'(I_0) (I-I_0)$ to first order, and using Fourier series, a rather general expression (see Appendix \ref{appendix_A} for details) can be derived:
 \begin{equation}
     \langle f \rangle (t)= \sum_{k=-\infty}^{\infty} f_k e^{-\frac{k^2 \sigma^2}{2}(\omega'(I_0)^2t ^2+1)} e^{-i k \omega(I_0) t}.
 \end{equation}
 Here $f_k=\int_{-\pi}^{\pi} f(\theta) e^{i k \theta} d\theta/(2 \pi)$ are Fourier coefficients. Thus we see the approach of the averages to $f_0$ as $t \rightarrow \infty$.  \textcolor{black}{The exponential time dependence of the numerical fit in Eq.~\ref{eq:regular_fit} is also present here.} The time scale for the equilibration is given by $t_* \sim 1/\sigma \omega'(I_0)$. With $\sigma=1/\sqrt{N}$, we get the relevant time scale of $\sqrt{N}/\omega'(I_0)$.
 The high frequency oscillations that accompany the 
 saturation are also captured fully by the classical expression and we see from here that they are related to the frequency $\omega(I_0)$ of the underlying orbit and their harmonics.
 For a concrete illustration, we may take $f(\theta)=\cos \theta$, so that the above procedure yields
 \begin{equation}
 \label{eq:varf_cos_integ}
 \begin{split}
     \text{var}f(t)&=\frac{1}{2}\left(1-e^{-\sigma^2 (\omega'(I_0)^2 t^2+1)}\right) \times \\
     &\left(1-e^{-\sigma^2 (\omega'(I_0)^2 t^2+1)} \cos[2  \omega(I_0)t]\right)
     \end{split}
 \end{equation}
 Thus the way these classical variances approach equilibrium is directly relevant to the saturation of quantum entanglement.

\begin{figure}
\includegraphics[width=1.0\linewidth, height=0.8\linewidth]{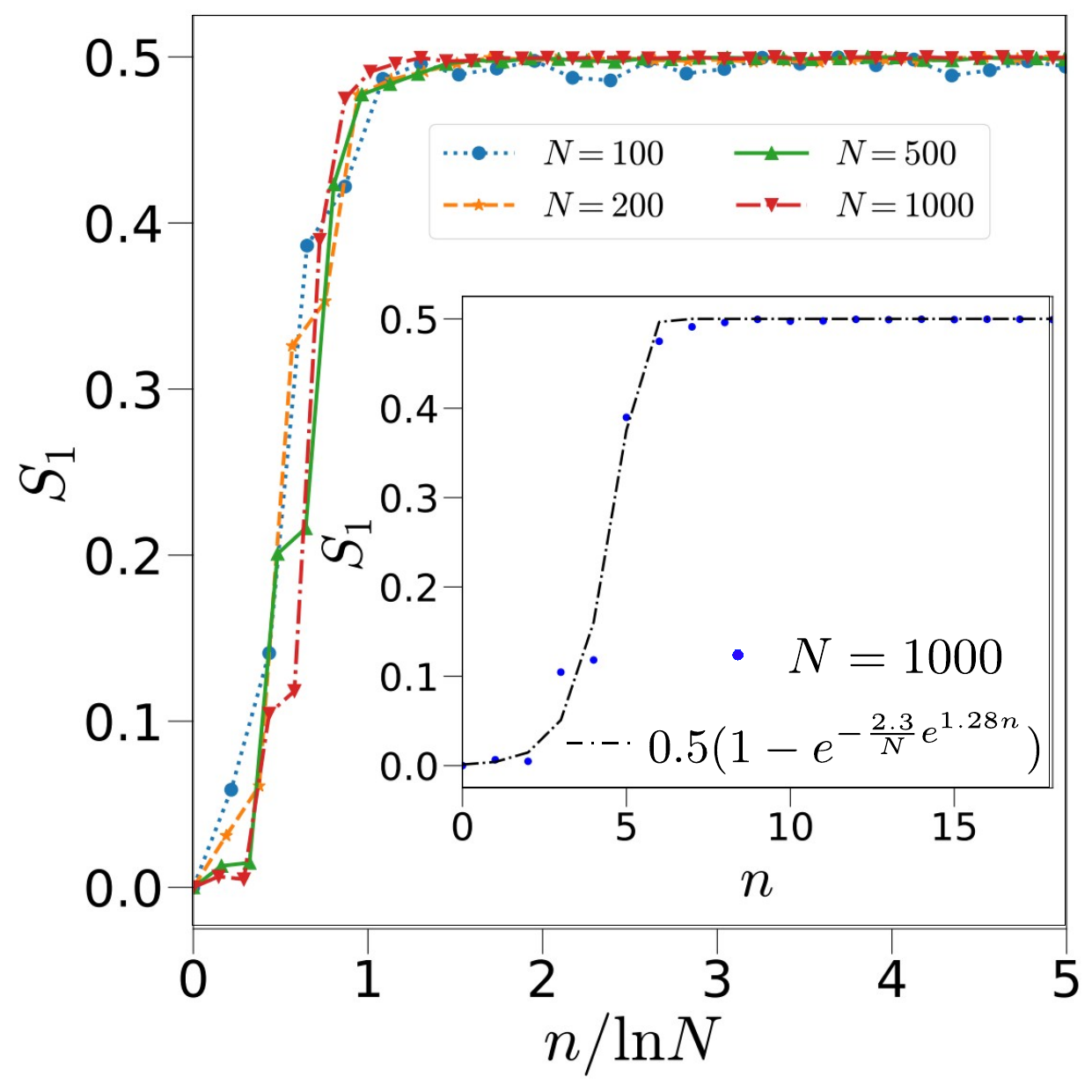}
\caption{The short time dynamics at $k=6$ for the chaotic case where Ehrenfest time goes as $\ln N$. Clearly, $S_1$ saturates when $n > \ln N$. There are no revivals in this case indicating chaotic dynamics. The inset shows comparison of analytical expression given in Eq.~\ref{eq:k6_disorderfree} with the numerics for $N=1000$.  The initial state is $|\theta,\phi\rangle= |2.25,1.1 \rangle$.}
\label{k6_disorderfree}
\end{figure}

We now discuss the  chaotic limit of the kicked top. Eq.~\ref{eq:ClassicalEnt} gives the saturation value $1/2$ easily enough, as any smooth ensemble of initial states tends to be uniformly distributed on the sphere and all the expectation values of $X_n$, $Y_n$ and $Z_n$ vanish. Thus the maximal entanglement of a single qubit corresponds to the mixing of a classical density over phase space. The form of the time evolution may also be inferred from general arguments to be approximately 
  \begin{equation}
  \label{eq:k6_disorderfree}
      \text{var}f(t) \approx \frac{1}{2}\left[ 1- \exp
      \left(-\alpha \sigma^2 e^{2 \lambda t}\right) \right],
  \end{equation}
where $\alpha>0$ is a constant, and $\lambda >0$ is the Lyapunov exponent. \textcolor{black}{The Ehrenfest time scale of chaotic systems, or the ``log-time", $\tau_{\hbar}=\ln N/(2 \lambda)$ emerges as the saturation time once we identify $\sigma^2=1/N$ .} See the Appendix \ref{appB} for a derivation of the variance for the exactly solvable cat maps and motivation for the form assumed above. The short time dynamics of $S_1$ in the chaotic limit $(k=6)$ is depicted in Fig.~\ref{k6_disorderfree} with the inset showing the proposed analytical form in Eq.~\ref{eq:k6_disorderfree}. We also note the absence of revivals in this case.

\subsubsection{Long time averaged linear entropy}
\label{longtimeaverage}

\begin{figure}[ht] 
\graphicspath{{old_images}}
  \label{ fig7} 
  \begin{minipage}[b]{0.5\linewidth}
    \includegraphics[width=\linewidth]{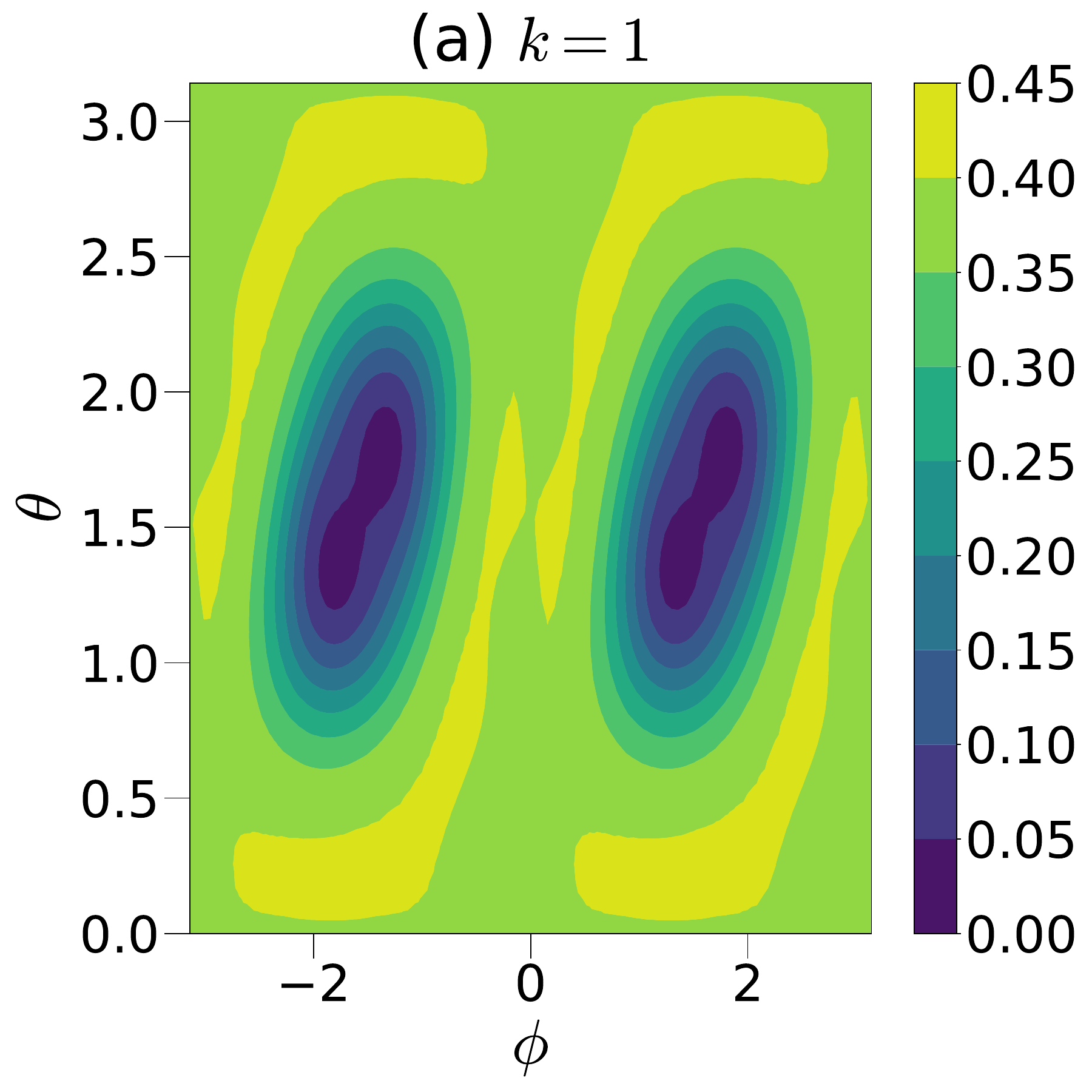} 
  \end{minipage} 
  \begin{minipage}[b]{0.5\linewidth}
    \includegraphics[width=\linewidth]{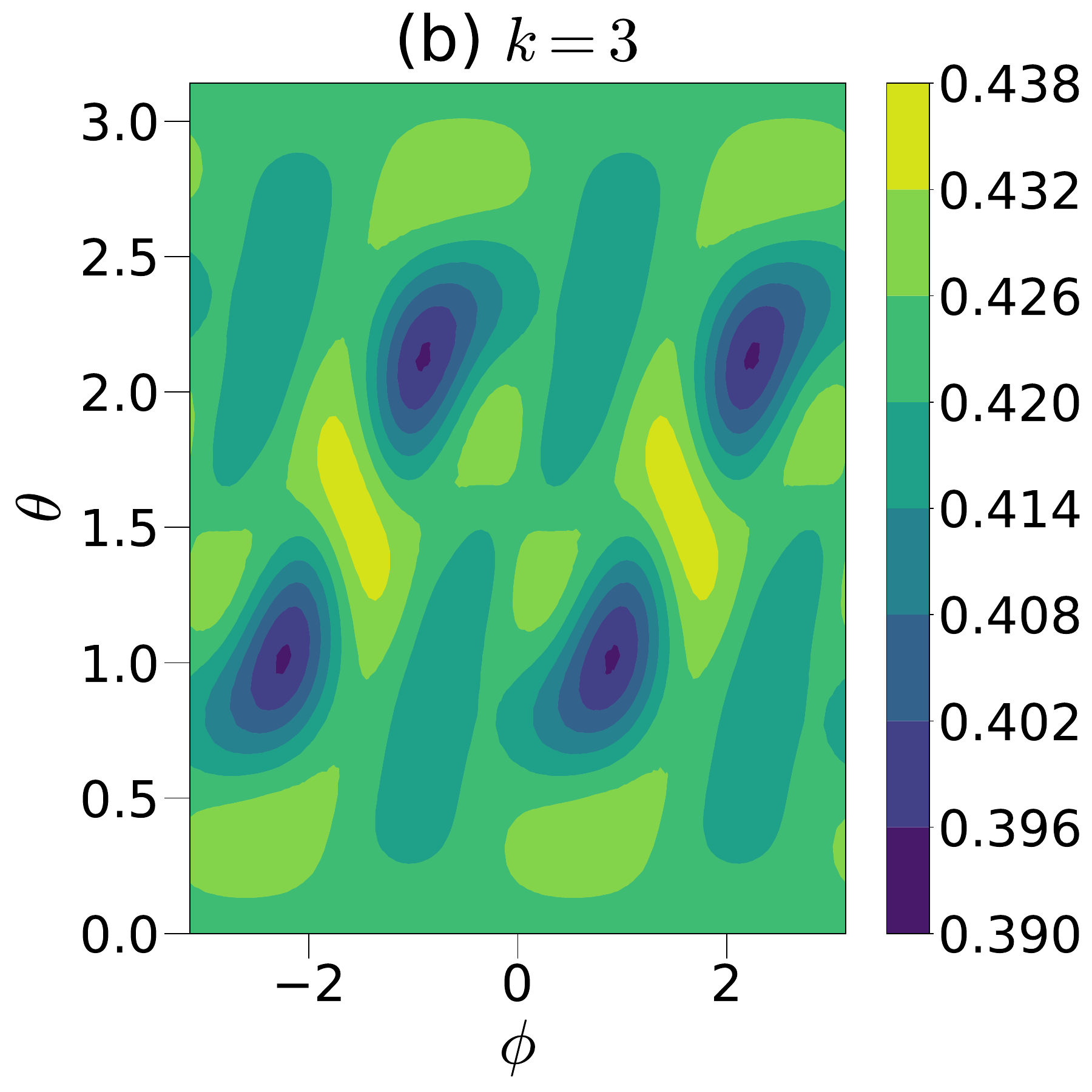} 
  \end{minipage}%% 
  \begin{minipage}[b]{0.5\linewidth}
    \includegraphics[width=\linewidth]{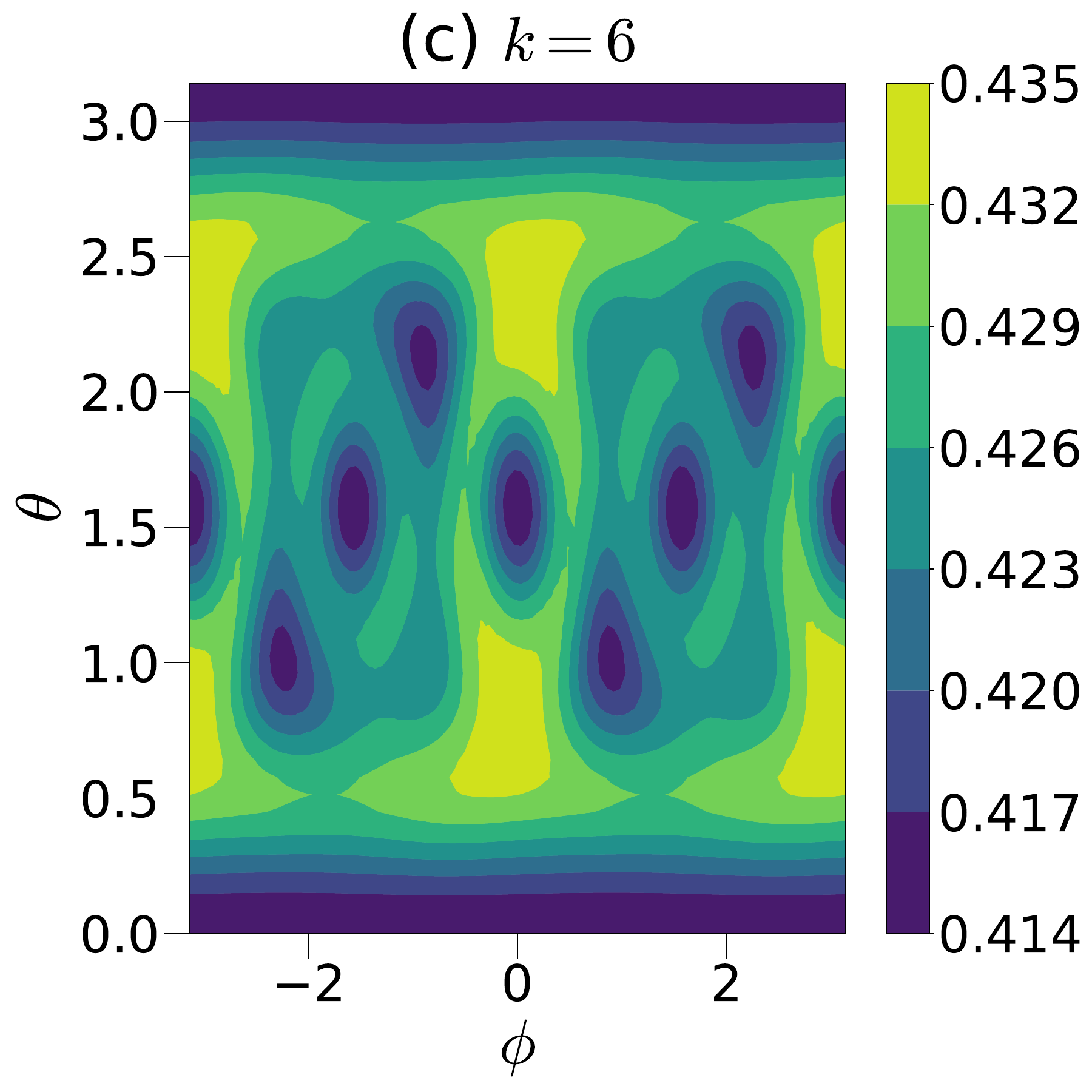} 
  \end{minipage} 
   \caption{(a), (b) and (c) shows the time averaged linear entropy plots (averaged upto 5000 kicks) for chaos parameters $k=1,3$ and $6$ respectively, by taking 5000 initial values of $(\theta,\phi)$ from the interval $\theta$  $\in$  $(0,\pi)$ and $\phi$  $\in$  $(-\pi,\pi)$.  Here we set $N=8$ qubits to compare with the disordered system. }
   \label{quantum_phase_space}
\end{figure}
Fig.~\ref{quantum_phase_space} shows the long time averaged linear entropy calculated for different initial states $|\theta, \phi \rangle$, and different $k$ values.
We see localized structures in Fig.~\ref{quantum_phase_space}(a) resembling classical phase space for $k=1$ around the fixed points $(\pi/2, \pi/2)$ and $(\pi/2,-\pi/2)$ where the linear entropy is minimum. A similar resemblance is seen for $k=3$ as well, see Fig.~\ref{quantum_phase_space}(b). %However, the structures appear different for $k=6$ as shown in  Fig. \ref{quantum_phase_space}.
For $k=6$ shown in Fig.~\ref{quantum_phase_space}(c), the linear entropy on an average gives $0.43$ consistent with earlier results of random permutation symmetric states \cite{tripartitescrambling}.

For the rest of the paper, in order to highlight initial state dependence of the dynamics after introducing the disorder, we select the following $|\theta, \phi\rangle$ for different $k$ values, and restrict the discussion to only these values.

\begin{enumerate}[label=(\alph*)]
\item For $k=1$, $|\theta=2.25, \phi=1.1 \rangle$ belonging to predominantly regular region
\item For $k=3$ regular, $|\theta=2.25, \phi=2.5 \rangle$ belonging to the regular region of mixed phase space.
\item For $k=3$ chaotic, $|\theta=2.25, \phi=1.1 \rangle$ belonging to the chaotic region of mixed phase space.
\item For $k=6$,  $|\theta=2.25, \phi=1.1 \rangle$ belonging to predominantly chaotic region.
\end{enumerate}

\section{Disordered kicked top: Quantum Dynamics}
\label{sec_disorder_interaction}
We now introduce disorder in the interaction term of the Hamiltonian thereby breaking the permutation symmetry of the system so that the dynamics can in principle explore the entire Hilbert space of $2^{N}$ dimensions  rather than the $N+1$ dimensional symmetric subspace. 
The Floquet operator $U'$ in the presence of disorder is given by:
\begin{equation}
U'=\exp\left ({-i\frac{k}{2N}\sum_{l< l'=1}^{N}(1+\epsilon_{ll'})\sigma_{l}^{x}\sigma_{l'}^{x} }\right) \exp\left ({-i\frac{\pi}{4}\sum_{l=1}^{N}\sigma_{l}^{y}}\right ),\\
\label{eq_disfloquet}
\end{equation}
\noindent where, as mentioned before, $\epsilon_{ll'}$ is a random number taken from a normal distribution with zero mean and standard deviation $w$.

The total angular momentum operator in presence of disorder is not a constant of motion, and its evolution has been discussed in Ref. \cite{manju24}. Below, we look at three different measures to understand the effect of the disorder on the dynamics, namely behavior of time scales, overlap with permutation symmetric basis, and change in effective dimension of the system. We will be mainly using linear entropy for these studies where $\langle S_j\rangle_w$ refers to disorder averaged linear entropy upto $j-$ spins whereas $\langle \overline{S_j}\rangle_w$ corresponds to disorder and long time averaged linear entropy. We also study behavior of $\langle \overline{S_j}\rangle_w$
for different initial states defined by $\theta, \phi$, and compare them with the classical phase space structures, which are well-defined only in the disorder free case. The disorder average is taken over 100 realizations and time average is taken from $10^5$ to $3 \times 10^5$ unless specified explicitly. We use fast Walsh Hadamard transform (FWHT) to go to system sizes upto $N=16$ \cite{fwht}.

\subsection{Effect on the Ehrenfest time scale and revivals}
Similar to the disorder free case, we examine the Ehrenfest and revival time scales, if they are present, focusing on $k=1$, which in the disorder free classical limit corresponds to regular dynamics.  In the absence of a classical limit with the disorder, we consider the Ehrenfest time scale as the saturation time for the single qubit linear entropy. It is to be noted that the system sizes that one can study in presence of disorder is much smaller than the sizes considered in the disorder free system.

\begin{figure}[H]
\includegraphics[width=1.0\linewidth, height=0.8\linewidth]{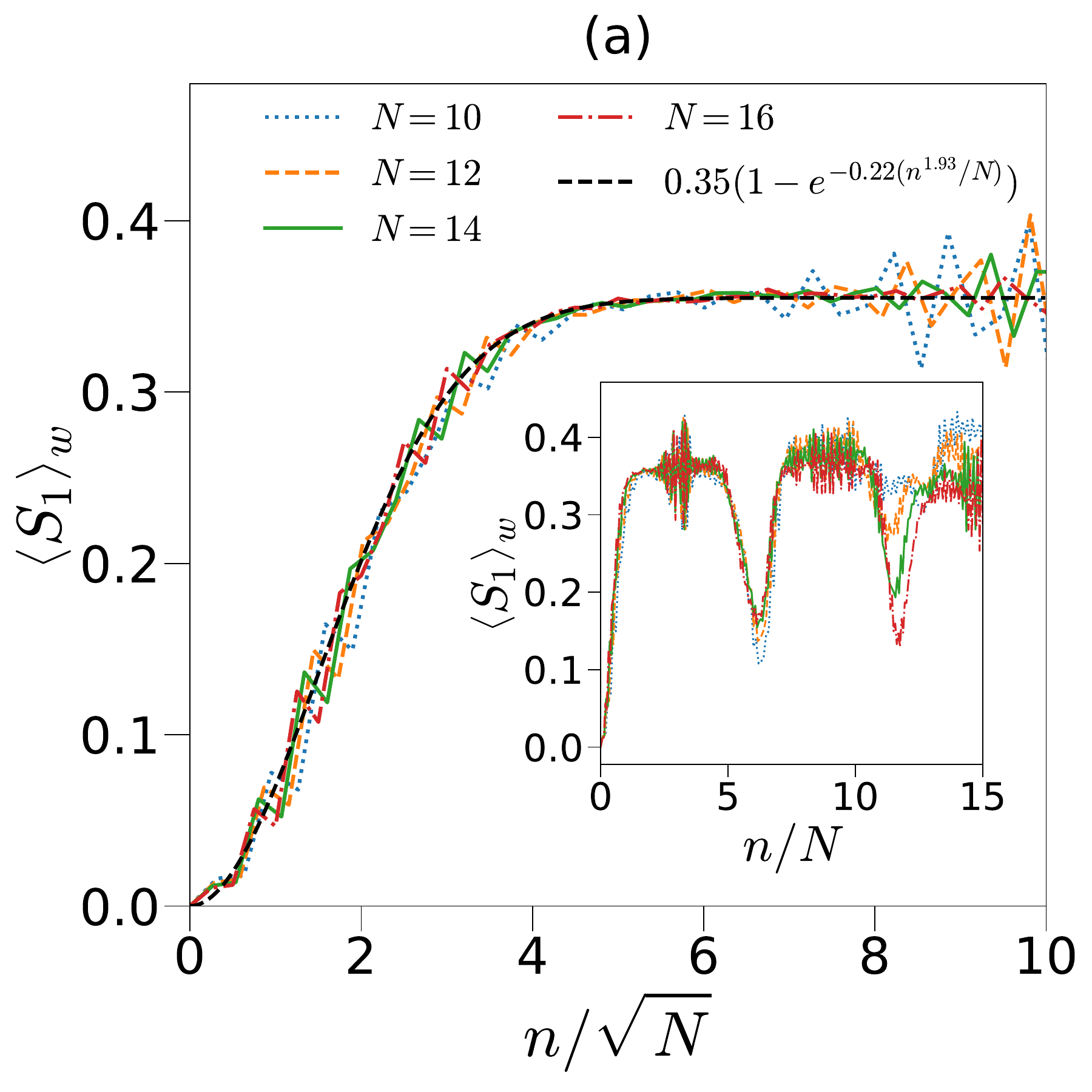}
\includegraphics[width=1.0\linewidth, height=0.8\linewidth]{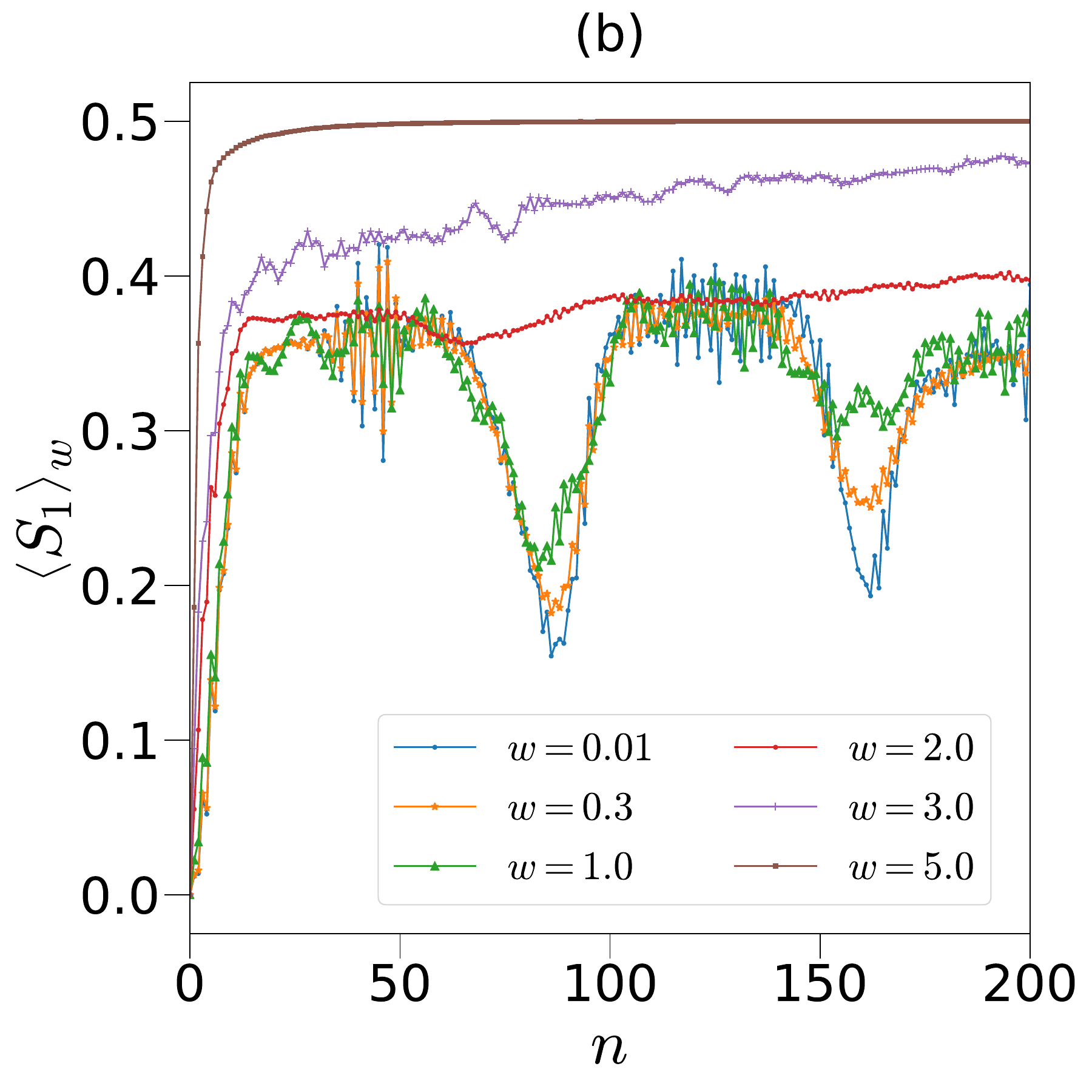}
\includegraphics[width=1.0\linewidth, height=0.8\linewidth]{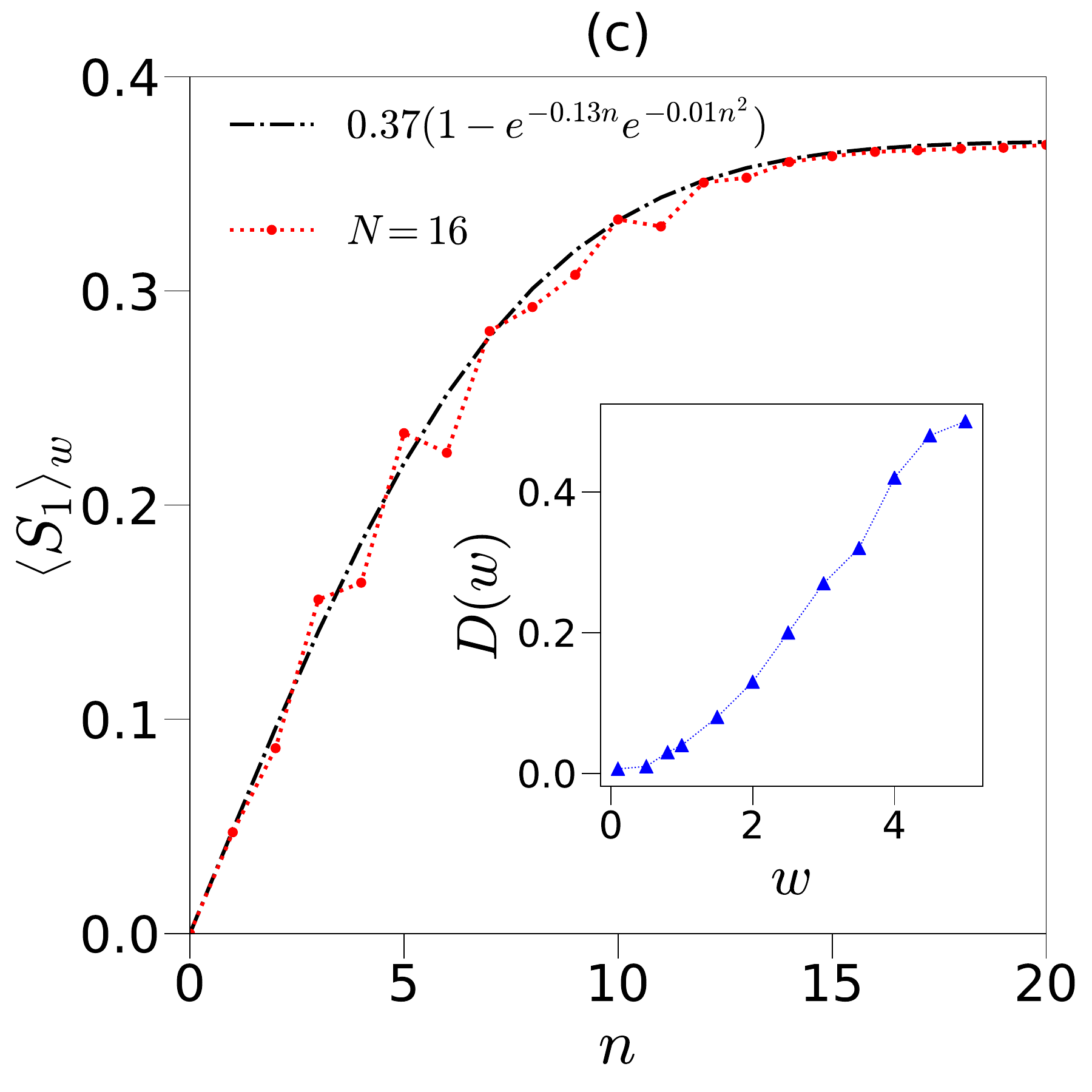}
 \caption{(a). The short time dynamics of linear entropy as a function of $n/\sqrt{N}$ for small disorder ($w=0.01$) showing $\tau_h \sim \sqrt{N}$. The inset shows $\left\langle S_1 \right \rangle_w$ with respect to  $n/N$ for larger times which clearly shows $t_r \sim N$ (b). Evolution of $\left\langle S_1 \right \rangle_w$ at different disorder strengths $w$ for $N=14$ demonstrating the gradual disappearance of revivals as $w$ is increased beyond $2$. (c). A comparison between numerics and the analytics for $w=2$, $N=16$. The inset shows the increase in $D$ with disorder $w$ for the same system parameters. All the figures correspond to $k=1$ with initial state set to $|\theta,\phi \rangle =|2.25,1.1 \rangle$. }
  \label{short_time_0.01var}
\end{figure}

Fig.~\ref{short_time_0.01var}(a) corresponds to a small disorder strength of $w=0.01$  ($w \ll 1$). In this case, the Ehrenfest time $\tau_{h}$ still scales as $\sqrt{N}$, similar to the disorder free case, as shown in the main figure. Inset of (a) highlights the revivals which scale as $N$ at least for the initial few revivals. Thus the delicate interference required for long-time quantum revivals survives small disorder, at least for initial states that are in the symmetric subspace.

 As the disorder increases we find a gradual destruction of the revivals as shown in Fig.~\ref{short_time_0.01var}(b) hinting at a chaotic phase in the large disorder limit. When $w \approx 2$, the revivals are destroyed, but the saturation value of the average single qubit  entanglement is approximately the same as for the ordered case's initial saturation. Beyond this value, the entanglement tends to approach the random state value of $\approx 1/2$, and this is reached rather quickly in the case when $w=5$, as shown in Fig.~\ref{short_time_0.01var}(b). 
 The entanglement entropy now saturates rapidly in comparison to the small disorder case. The growth now seems to be similar to the $k=6$ clean model shown in Fig.~\ref{k6_disorderfree}. However, there are important differences and in fact the growth appears to be linear before saturation. The lack of a classical model and Lyapunov exponent in this disordered system changes qualitatively the secular behavior, although the saturation in both cases is nearly maximal at $\approx 0.5$.

 As a plausible heuristic mechanism to understand the differences, we may model the effect of the disorder as a noise, since the reduced density matrix evolution of a single qubit is going to be given by some completely positive trace preserving (CPTP) map. As the classical density evolutions were insightful in the disorder free cases, here we can model the disorder as leading to an effective noisy classical dynamics. The noisy kicked top will not preserve the dynamics on a unit sphere and much like in the quantum case will diffuse away from this.

 We extend the disorder free integrable system calculations by introducing noise, i.e; $\theta(t)=\theta(0)+\omega(I)t+\eta(t)$ where $\eta(t)$ is the cumulative noise with probability distribution \\
 \[P(\eta)=\frac{1}{\sqrt{2\pi Dt}}e^{-\frac{\eta^{2}}{2Dt}} \] with $D$ being the diffusion constant.
 In Appendix~\ref{AppC} we derive the equivalent of the example of $f(\theta)=\cos \theta$ of Eq.~\ref{eq:varf_cos_integ}, which is given by
 \begin{equation}
 \label{eq:varf_integ_cos_noise}
\begin{split}
&\mbox{var}\,f(t)=\frac{1}{2}(1-e^{-Dt}e^{-\sigma^2(1+\omega'(I_0)^2t^2)}) \\ & \times (1-e^{-Dt}e^{\sigma^2(1+\omega'(I_0)^2t^2)}\cos[2\omega(I_0)t]).
\end{split}
\end{equation}

Thus, the initial quadratic growth in the disorder free case is replaced by a linear one where the diffusion constant $D$ will grow with the disorder $w$. A comparison between the numerical calculations and an approximation of the analytical expression of Eq.~\ref{eq:varf_integ_cos_noise} given by $S_0(1-e^{-Dn}e^{-\alpha n^2})$ is shown in Fig.~\ref{short_time_0.01var}(c) for $w=2$, with the inset illustrating the variation of the diffusion constant $D$ with $w$.
\textcolor{black}{It maybe noted that in open quantum systems, with or without quantum chaos, a proxy for the environment is noise in the system, for example see \cite{Bidhi2025, quanum_ratchet_PRL, dissipative_chaos_tavis, chaos_magic_QKT}, and our treatment can be understood from such a perspective.}
  
\textcolor{black}{In summary, for the case of $k = 1$ the revivals seen in the disorder free limit are robust against small disorder. On the other hand, as the disorder is increased, the revivals disappear similar to the case of disorder-free chaotic system. This hints towards the generation of random dynamics even for small values of $k$ as the disorder is increased.}
While this random phase has similarities to the disorder free chaotic case, there are important differences. To characterize the new random/chaotic phase, we study other quantities below.

\subsection{Overlap with permutation symmetric basis}
For quantifying how much the system remains in the permutation symmetric subspace (PSS), we define $\chi$ as the sum of the square of the overlap of the evolved state in the full $2^{N}$ basis with each of the $(N+1)$ permutation symmetric or Dicke basis \cite{tripartitescrambling, Haakebook, russomanno2021quantum}. The overlap is $1$ for the disorder free case and $0$ if the system completely goes out of the symmetric subspace.\\
\begin{comment}
We explain below the calculation of $\chi$ using an example of a 3 qubit state. Let $\ket{\Psi^{n'}_{2^{3}}}$ is the evolved state after n time steps in the full $2^{3}$ basis which can be written as:
\begin{equation}
\begin{split}
\ket{\Psi^{n'}_{2^{3}}}=a_{0}\ket{000}+a_{1}'\ket{001}+a_{1}''\ket{010}+a_{2}'\ket{011}\\
+a_{1}'''\ket{100}+a_{2}''\ket{101}+a_{2}'''\ket{110}+a_{3}\ket{111}.
\end{split}
\end{equation}
The permutation symmetric basis set is given by:
\[\ket{000}=\ket{0_{3}},\]
\[\frac{\ket{001}+\ket{010}+\ket{100}}{\sqrt{3}}=\ket{1_{3}},\]
\[\frac{\ket{011}+\ket{110}+\ket{101}}{\sqrt{3}}=\ket{2_{3}},\]
\[\ket{111}=\ket{3_{3}}.\]
The overlap of $\ket{\Psi^{n'}_{2^{3}}}$ onto each of the permutation symmetric basis is given by:
   \[ \braket{0_{3}|{\Psi^{n'}_{2^{3}}}}  =a_{0}= \alpha_{1},\]
   \[ \braket{1_{3}|{\Psi^{n'}_{2^{3}}}} =\frac{a_{1}'+a_{1}''+a_{1}'''}{\sqrt{3}}=\alpha_{2},\]
   \[ \braket{2_{3}|{\Psi^{n'}_{2^{3}}}} =\frac{a_{2}'+a_{2}''+a_{2}'''}{\sqrt{3}}=\alpha_{3} ,\]
   \[ \braket{3_{3}|{\Psi^{n'}_{2^{3}}}} =a_{3}=\alpha_{4}. \]
The quantity $\chi$ is then calculated as:
\begin{equation}
\chi = |\alpha_1|^{2}+|\alpha_2|^{2}+|\alpha_3|^{2}+|\alpha_4|^{2}.
\end{equation}

Clearly $\chi = 1$ when the dynamics is restricted to permutation symmetric subspace.
\end{comment}

\begin{figure}
\centering
  \includegraphics[width=1.0\linewidth, height=0.8\linewidth]{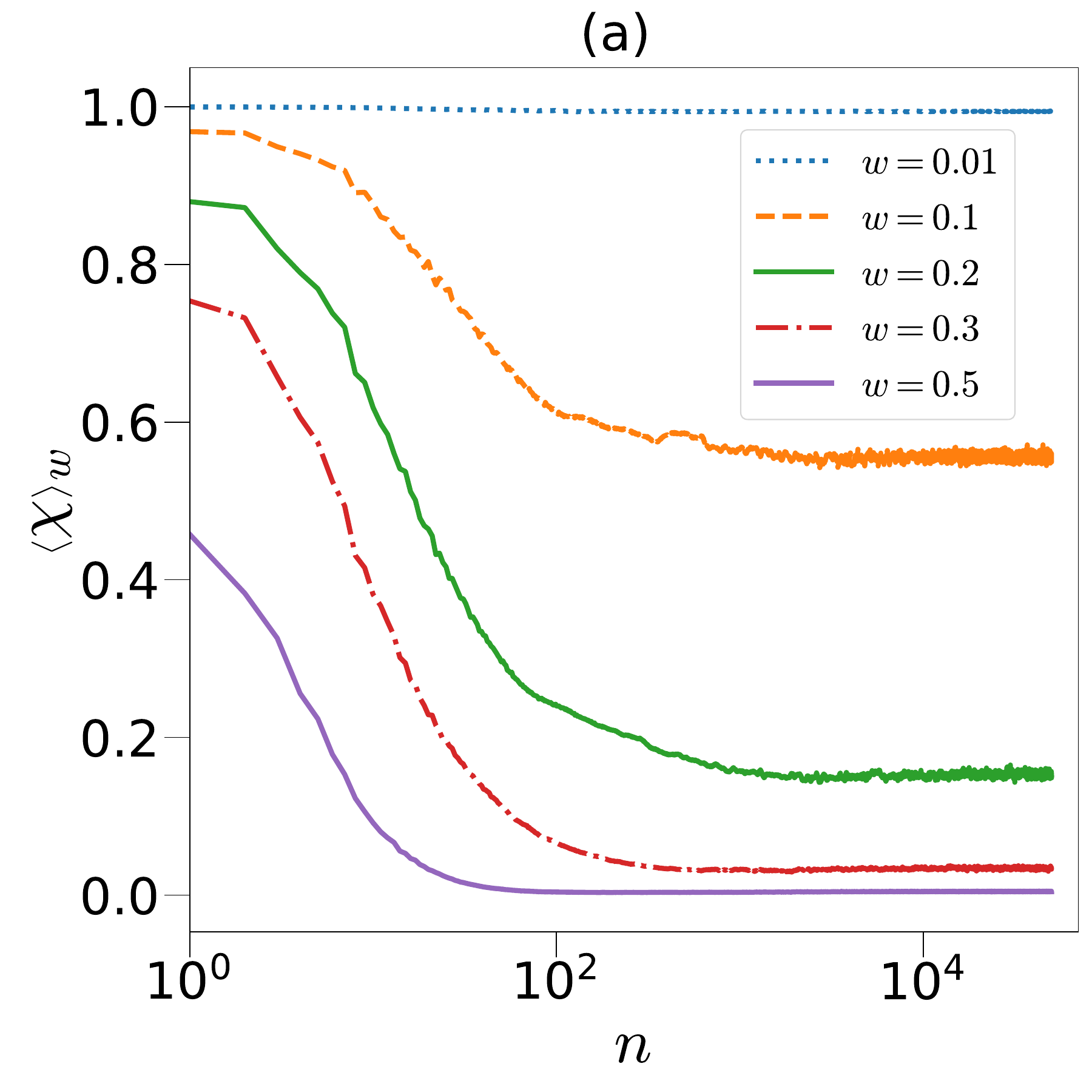}
 \includegraphics[width=1.0\linewidth, height=0.8\linewidth]{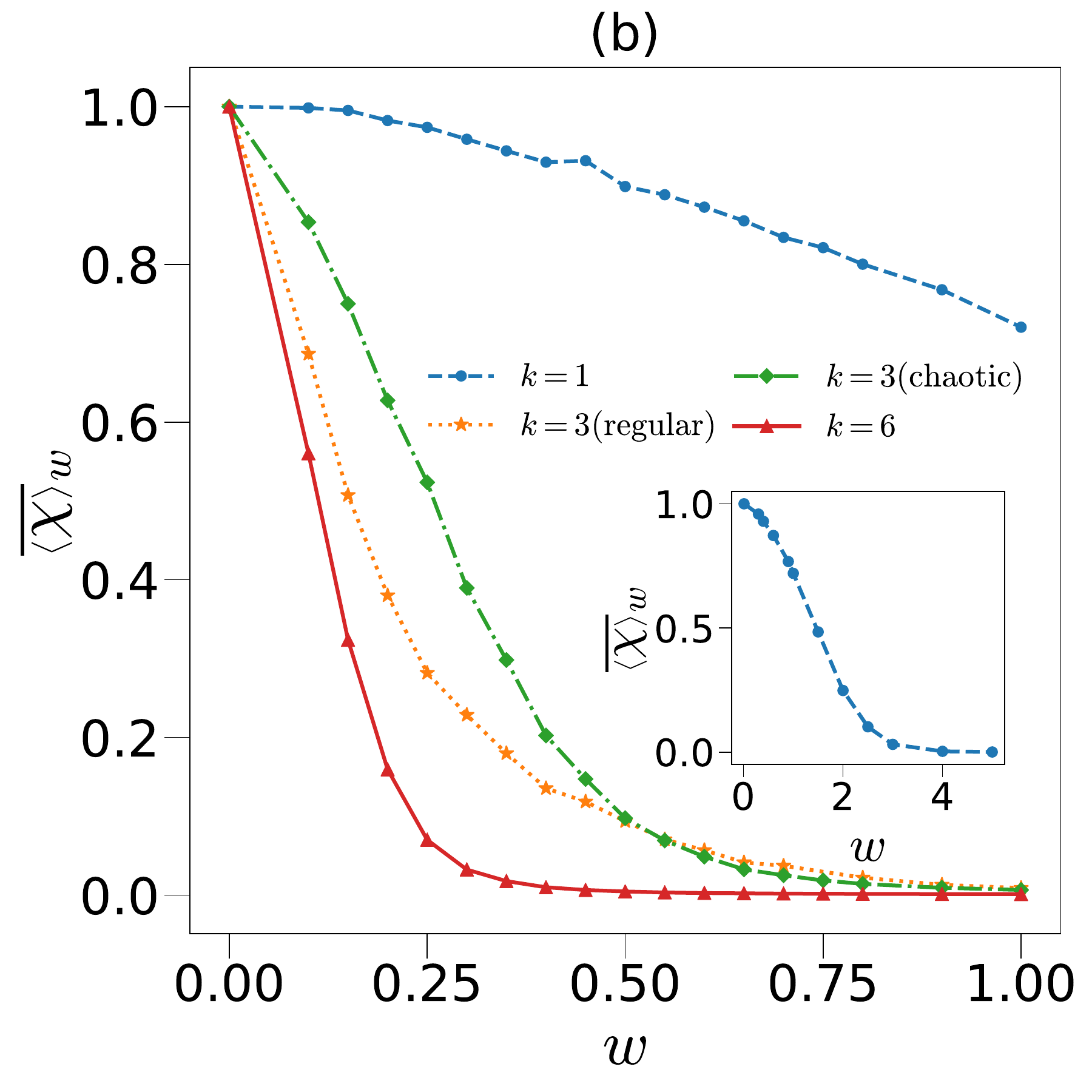}
 \caption{(a). The plot of $\left \langle \chi \right \rangle_w$ with respect to kicks $n$ (averaged over 100 disorder realisations) for different $w$ at $k=6$ and $N=14$. Initial state is $|2.25,1.1\rangle$. (b) The plot of time averaged $\chi$ (averaged over saturation values for 100 disorder realisations) vs $w$ for different chaos parameter $k$. The inset shows the case for $k=1$ for large disorder where the overlap $\chi$ eventually goes to zero. The initial states are as discussed in the text.}
 \label{figoverlap}
\end{figure}

The variation of disorder averaged overlap $\left \langle \chi \right \rangle_{w}$ as a function of time $n$ for different disorder strengths $w$ is shown in
Fig.~\ref{figoverlap}(a) for $k=6$. We find that for small $w$ of the order of $0.01$, the overlap is very close to 1 pointing to a dynamics which is essentially restricted within the PSS. The overlap is around ($\sim 60\% $) when $w$ is 0.1, whereas it decreases to nearly $0$ when $w$ is increased to $0.5$, signaling towards a dynamics for which the permutation symmetric subspace is not special.
Similar behavior is observed for $k=1$ and $k=3$ as well. 

\begin{figure}[h]
\centering
    \includegraphics[width=1.0\linewidth, height=0.8\linewidth]{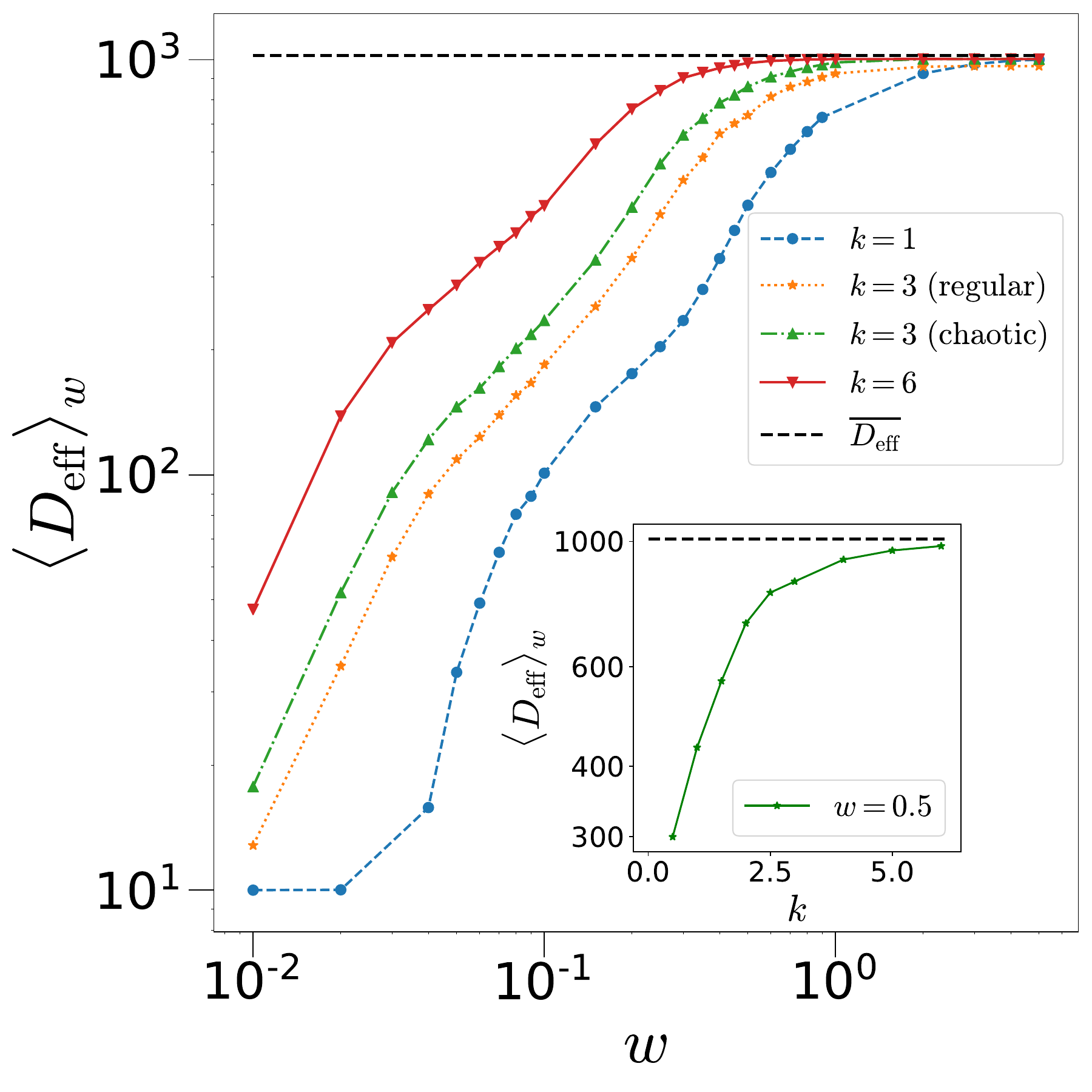}
 \caption{The plot of $\langle D_{\rm{eff}} \rangle_w$ with respect to disorder strength $w$ for different values of $k$ for $N=10$ qubits. For comparison, $D_{\rm{eff}} \sim 11$ for all $k-$values in the disorder free case. \textcolor{black}{Inset: Variation of $\langle D_{\rm {eff}} \rangle_w$ as a function of $k$ for $w=0.5$ and $N=10$. We see that with increase in $k$ values, $\langle D_{\rm eff} \rangle_w$ also increases and approaches $\overline{D_{\rm eff}}$ (indicated by dashed line) for large $k$. The initial state is $\ket{\theta,\phi}=\ket{2.25,1.1}$}.}
 \label{fig_eff_dim}
\end{figure}

Fig.~\ref{figoverlap}(b) shows the variation of time  and disorder averaged overlap $\overline{\left \langle \chi \right \rangle}_w$ as a function of $w$ for different $k$. It can be seen that the evolved state for $k=1$ is still mostly in the permutation symmetric subspace at least upto $w=1.0$ and approaches zero as $w$ is increased to 5, see inset of Fig.~\ref{figoverlap}(b). For $k=6$, it reaches almost zero for $w$ above 0.3. The $k=3$ case whose classical phase space shows mixed dynamics, lies intermediate between $k=1$ and $k=6$. The study of $\overline{\left \langle \chi \right \rangle}_w$  shows that indeed as $w$ is increased for a given $k$, the system evolves more and more out of the PSS.  Also, small $k$ requires larger disorder to be taken out of PSS whereas large $k$ can be taken out of PSS even by small disorder, once again consistent with our earlier results \cite{manju24}.

\subsection{Eigenstates and an Effective Dimension}
The previous sections have shown that for a given $k$, the evolution starts from $N+1$ dimensional PSS and is eventually taken out of this space as the disorder is increased where the dynamics seem to be chaotic and there are no revivals.
We now explore the complementary issue of how the eigenstates change as the disorder strength is increased.
To answer this question, we express a given spin coherent state which is a member of the PSS, in terms of Floquet eigen states $\ket{\phi_{i}}$ of Eq.~\ref{eq_disfloquet}:
\begin{equation}
    \ket{\theta \phi }= \sum_{i=1}^{2^N} c_{i}\ket{\phi_{i}}.
\end{equation}

The effective dimension $D_{\rm {eff}}$ is defined  as the minimum value of $K$ such that $\sum_{i=1}^{K}|c_{i}|^{2}=1-\alpha$, where the coefficients are arranged in decreasing order of their magnitude and $\alpha$ is a small number we choose \cite{Haakebook, haake1987classical}. \textcolor{black}{Thus, $D_{\rm{eff}}$ for a given state is a measure of it's delocalization in the Floquet eigenbasis. It is invariant under the time evolution of the state, hence, $D_{\rm{eff}}$ is also equal to the dimension of the Hilbert space that the system explores during the dynamics. Here, we have chosen $\alpha$ such that the sum of the intensities is $0.9999$, that is the deviation from normalization is about $.01\%$.} 
Figure~\ref{fig_eff_dim} shows the plot of disorder averaged $\langle D_{\rm {eff}} \rangle _w$ as a function of $w$ for $N=10$ and $\alpha=0.0001$, and the state $\ket{\theta \phi}$ is taken as discussed in \ref{longtimeaverage} for different $k$ values, {\textcolor{black}{whereas the inset shows the variation of $\langle D_{\rm {eff}} \rangle _w$ with $k$ for a fixed $w$}}. \textcolor{black}{The behavior of $\langle D_{\rm eff} \rangle_w$ for small enough $\alpha$ is monotonic with respect to the disorder strength $w$ for a fixed $k$. It is observed that as $w$ is increased, 
$\langle D_{\rm{eff}} \rangle_w$ increases from $N+1$ to nearly the maximum dimension of $2^{10}=1024$. In fact we may estimate the dependence of this saturation value on the parameter $\alpha$ using random states, which we denote as  $\overline{D_{\rm {eff}}}$. As shown in Appendix \ref{appDeff}, $\overline{D_{\rm {eff}}}$ is given by
\begin{equation}\label{eq:D}
\overline{D_{\rm {eff}}}\approx (1-\sqrt{2\alpha}+\alpha/3)\,2^N
\end{equation}
for $\alpha \ll 1$. This follows from applying the order statistics of a randomly broken interval to random quantum states \cite{Lak_extreme,David_orderstats}, where the intensities represent the length of the pieces of a unit interval.}

As also indicated by $\chi$, a measure for dynamics, the stationary states of the system are also essentially within the  PSS for small disorder whereas they explore nearly the full Hilbert space as $w$ is increased. Also, it reaches almost the full Hilbert space value for smaller $w$ when the chaos parameter $k$ is large.
It is noteworthy that the effective dimension only approach the maximum possible value for fairly large disorder strength. \textcolor{black}{In particular, a well localised coherent state on a periodic orbit/ fixed point of the clean model has $D_{\rm{eff}}$ that deviates significantly from the random state value. This deviation indicates that the corresponding eigenstates are scarred by this orbit. This can be an example of many-body scars with the difference that here the scars can be identified with a classical model which has been disordered. A more detailed study on this is left for future work.} In particular this indicates that eigenstates can be substantially scarred by certain PSS coherent states and will be an example of many-body scars with the difference that here the scars can be identified with a classical model which has been disordered.  A more detailed study on this is left for future work.

\subsection{Long-time averaged linear entropy}
\label{long_time_interaction}

\label{quantum_phase_space_disorder}
\begin{figure}[ht]
\centering
 \includegraphics[width=1.0\linewidth,height=0.8\linewidth]{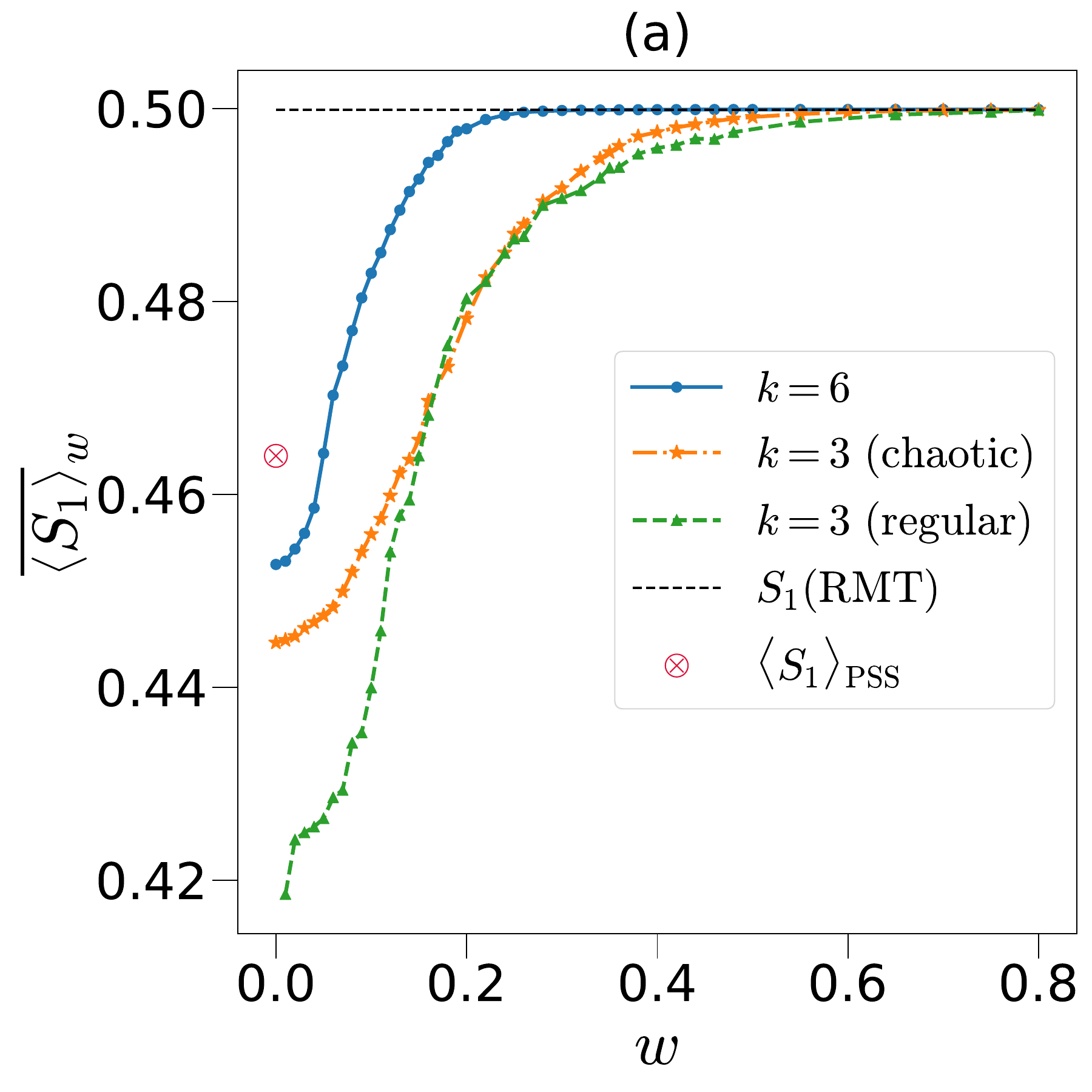}
 \includegraphics[width=1.0\linewidth,height=0.8\linewidth]{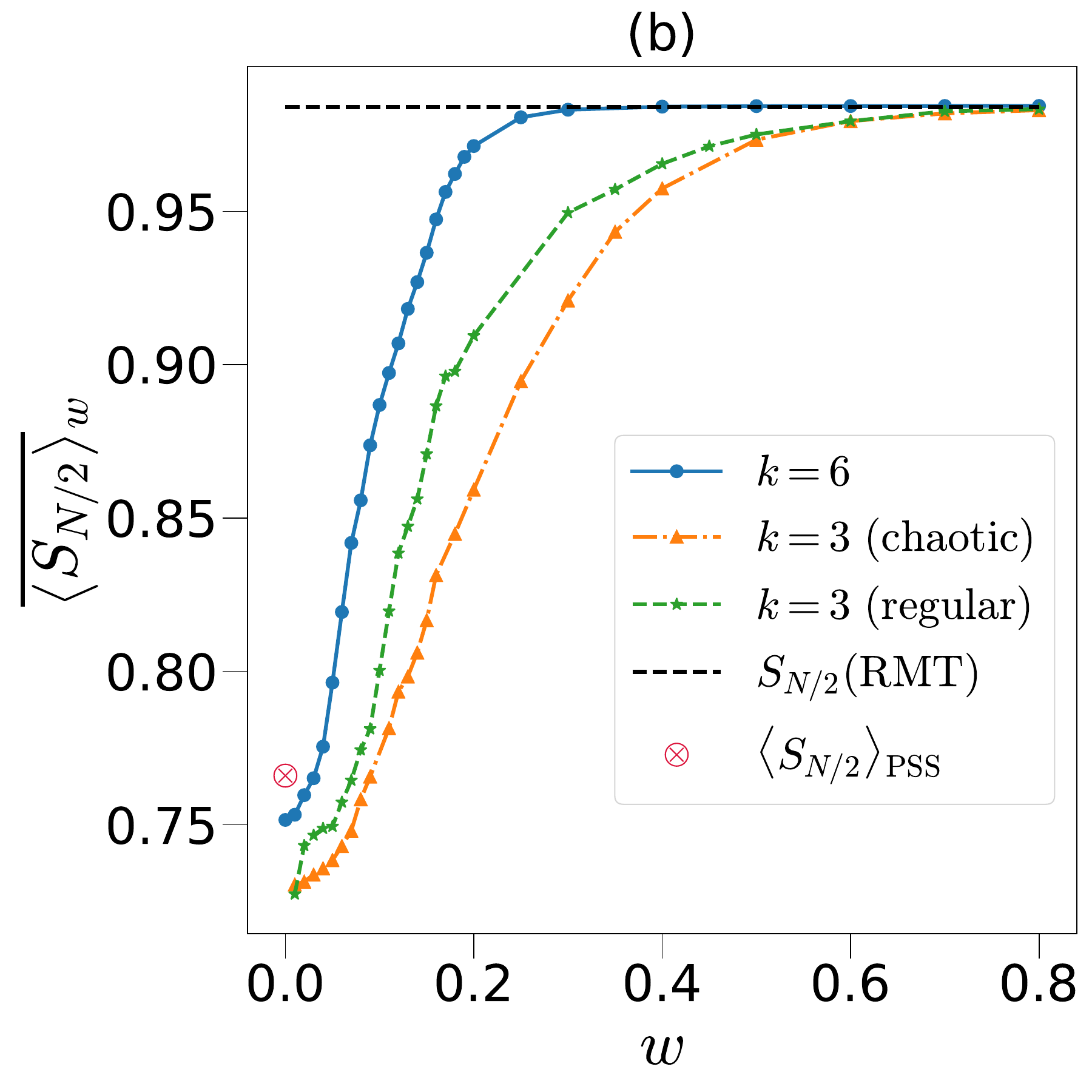}
  \caption{(a): The long time and disorder averaged linear entropy with respect to $w$ for $N=14$ qubits at different chaos parameter $k$ for $1:N-1$ bipartition.~(b): The plot for $N/2:N/2$ bipartition for the same parameters. The initial states are chosen as discussed in \ref{longtimeaverage}.  }
  \label{long_time_disorder_int}
\end{figure}

\begin{figure*}[ht]
    \graphicspath{phase_space_disorder}
    \includegraphics[width=0.3\linewidth]{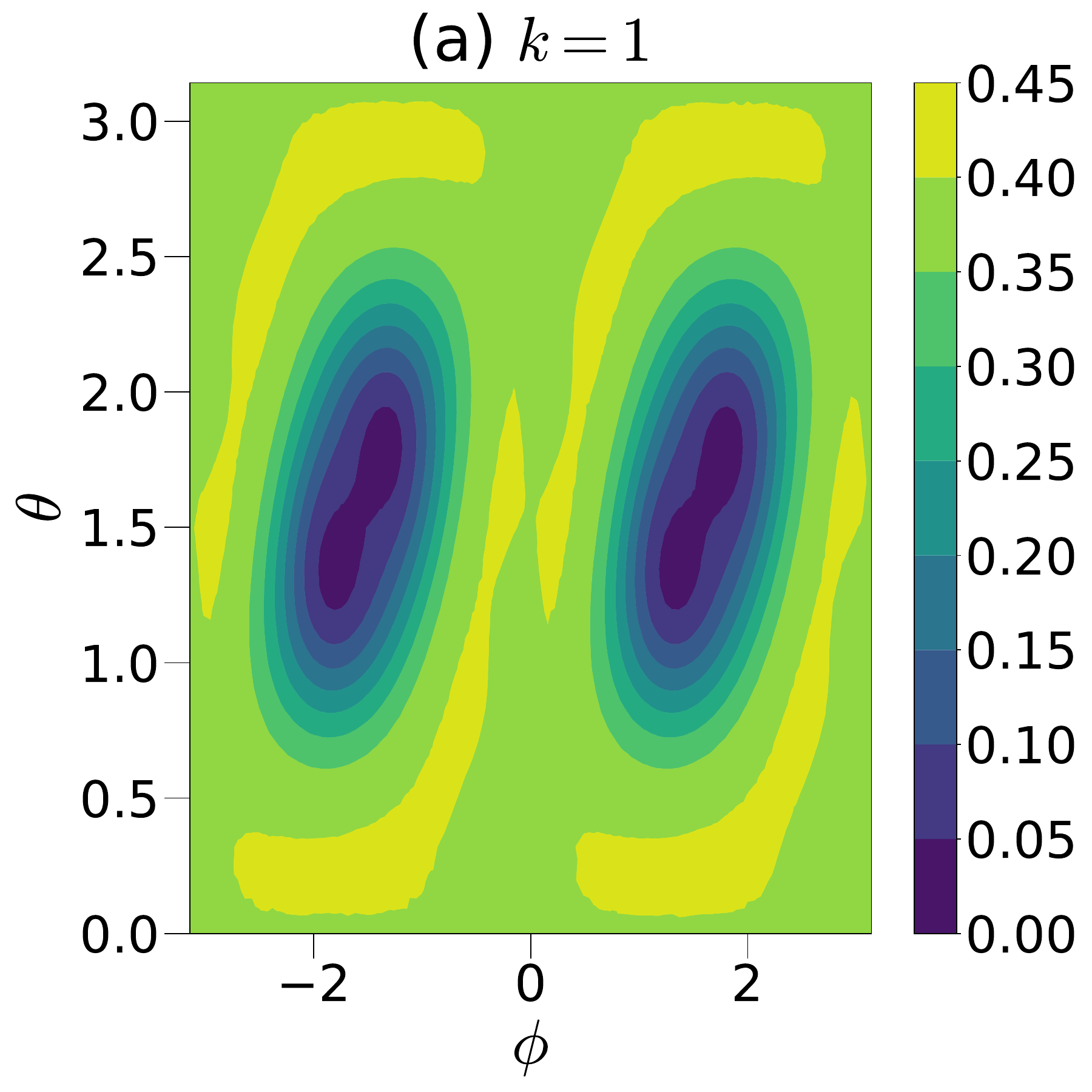}%
    \includegraphics[width=0.3\linewidth]{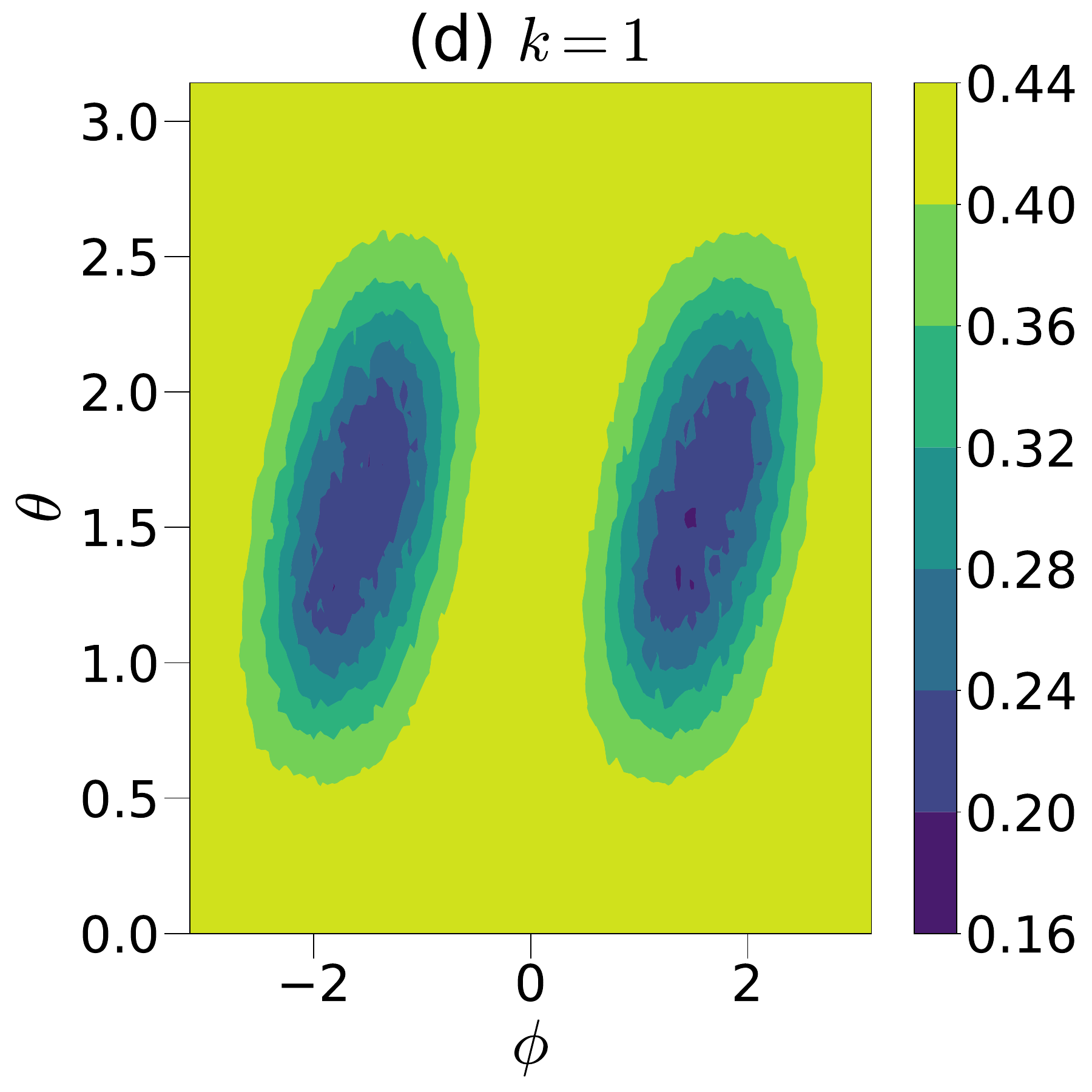}%
     \includegraphics[width=0.3\linewidth]{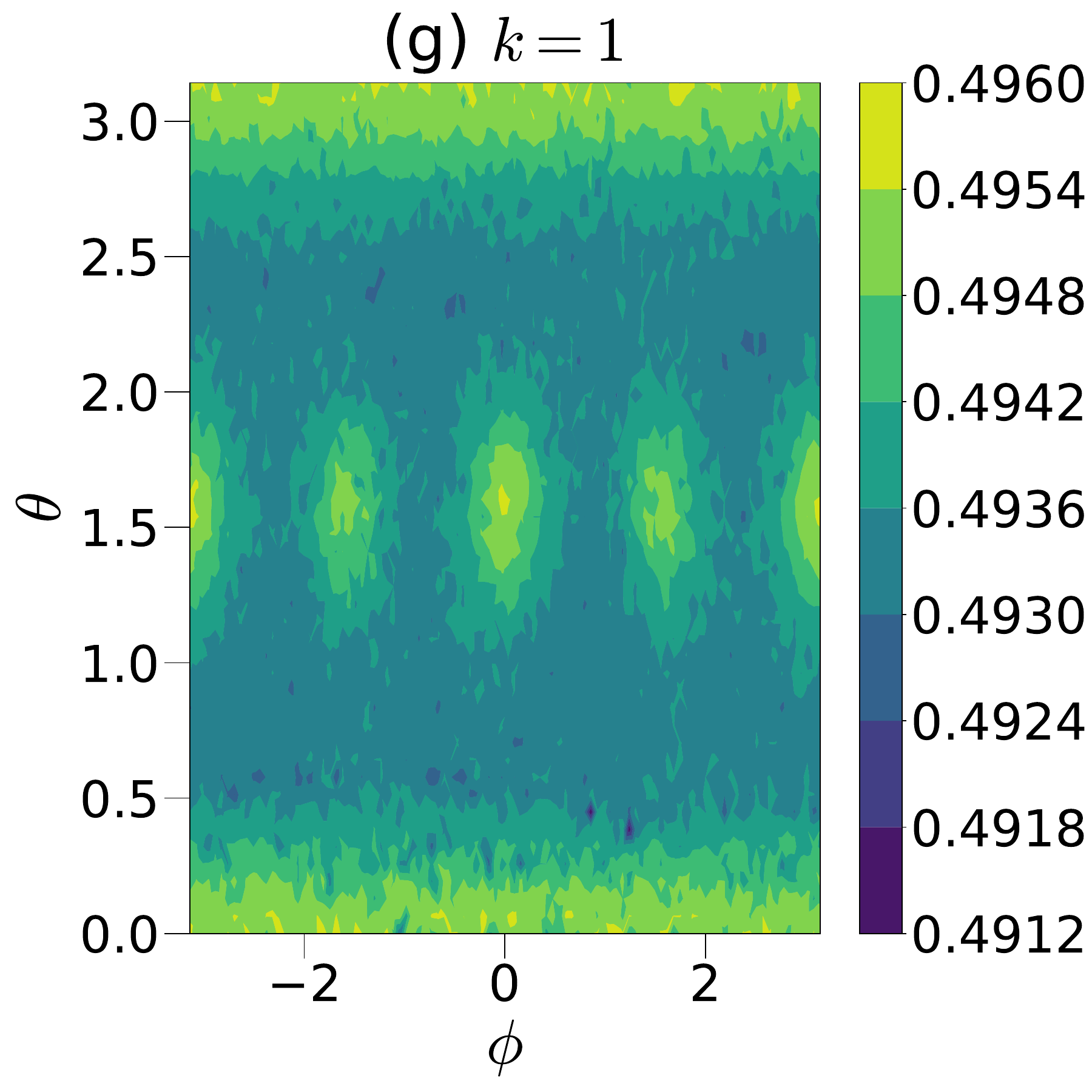}
     \includegraphics[width=0.3\linewidth]{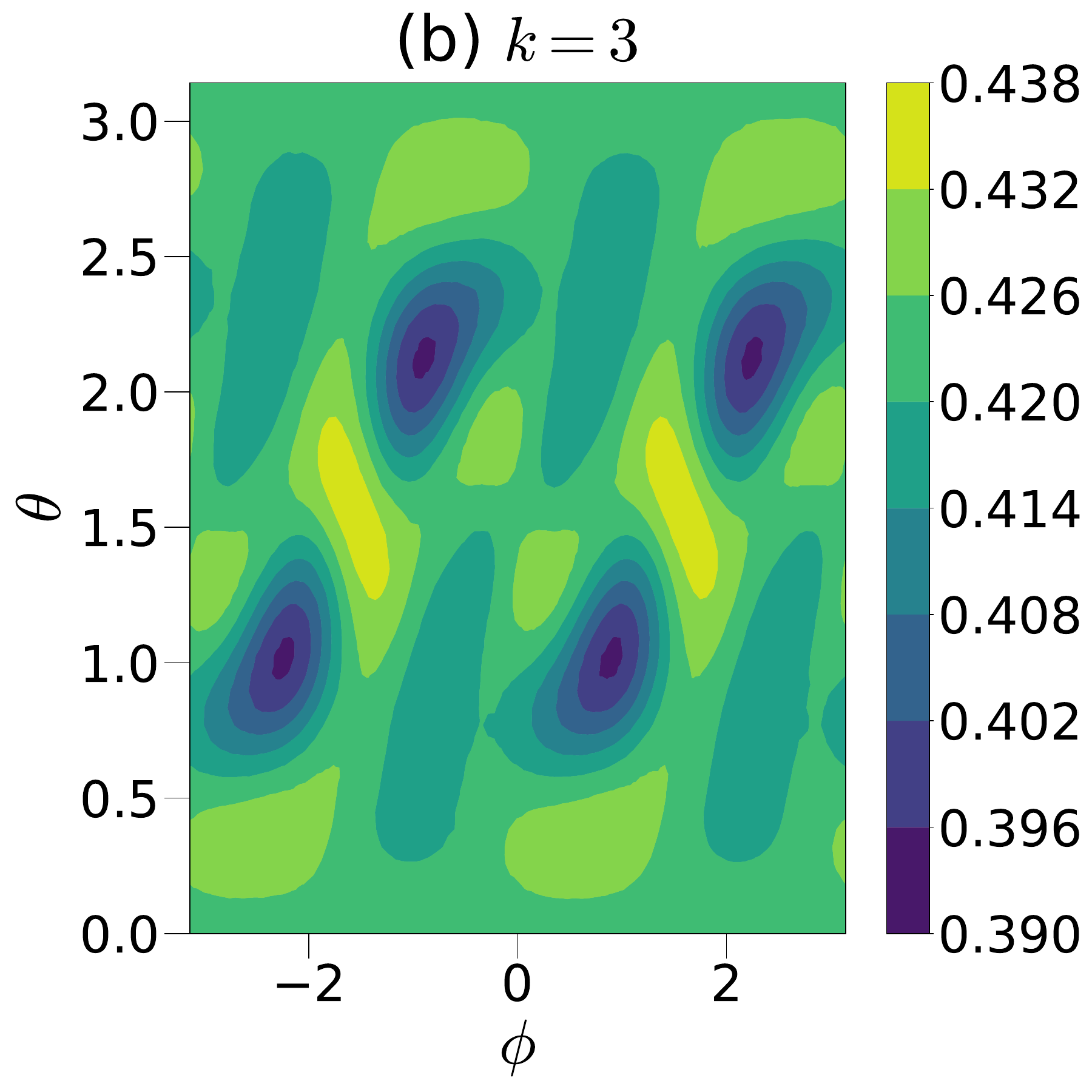}%
    \includegraphics[width=0.3\linewidth]{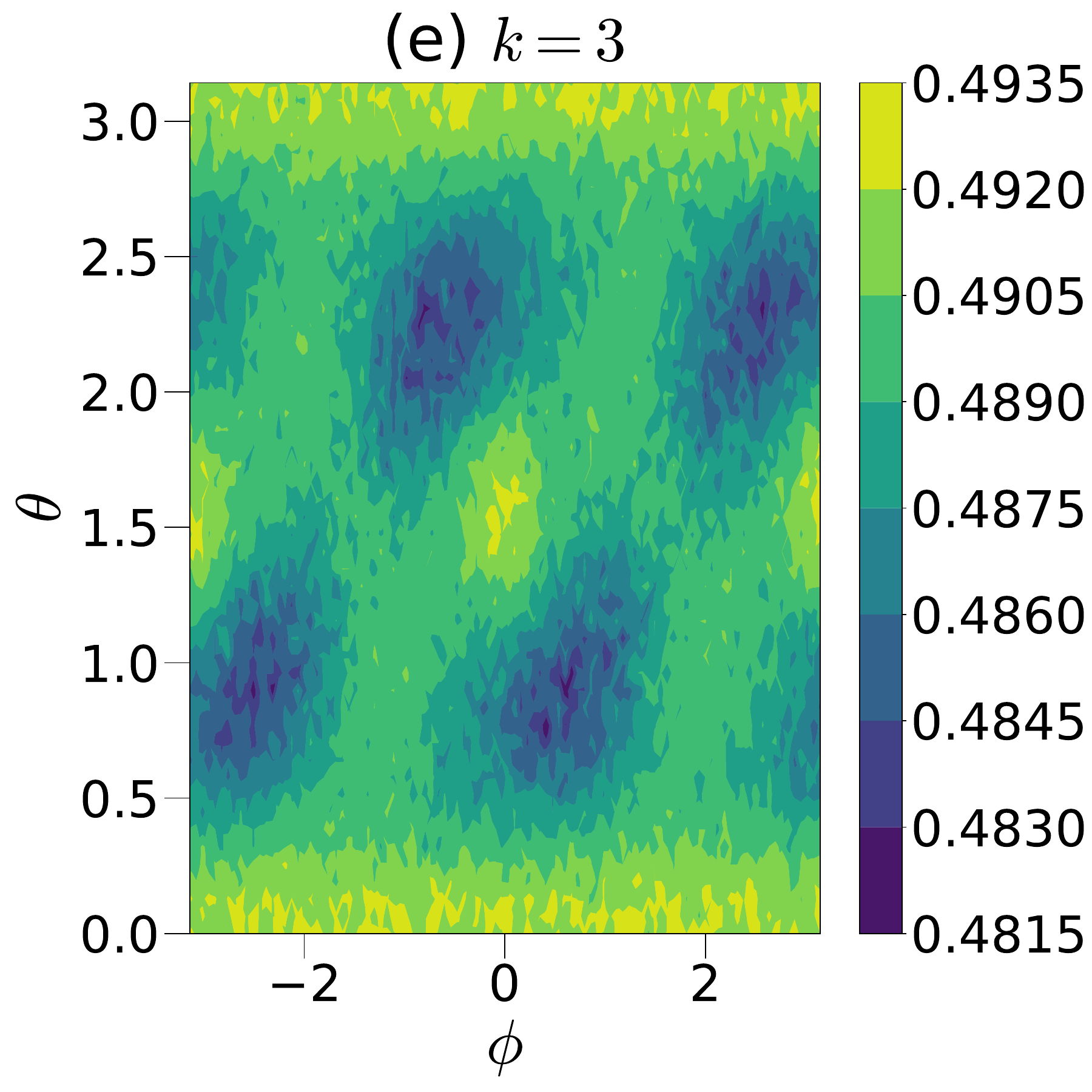}%
      \includegraphics[width=0.3\linewidth]{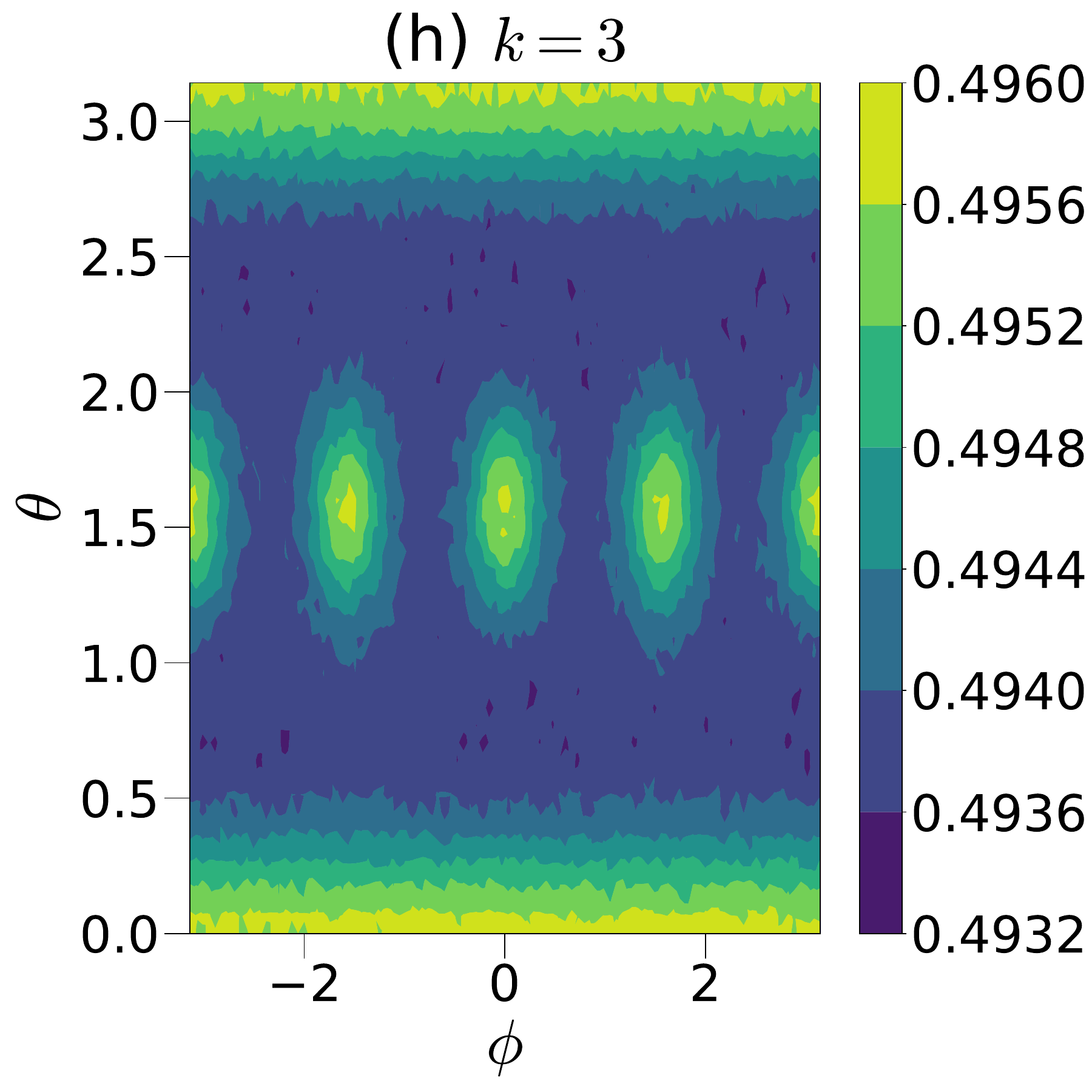}
     \includegraphics[width=0.3\linewidth]{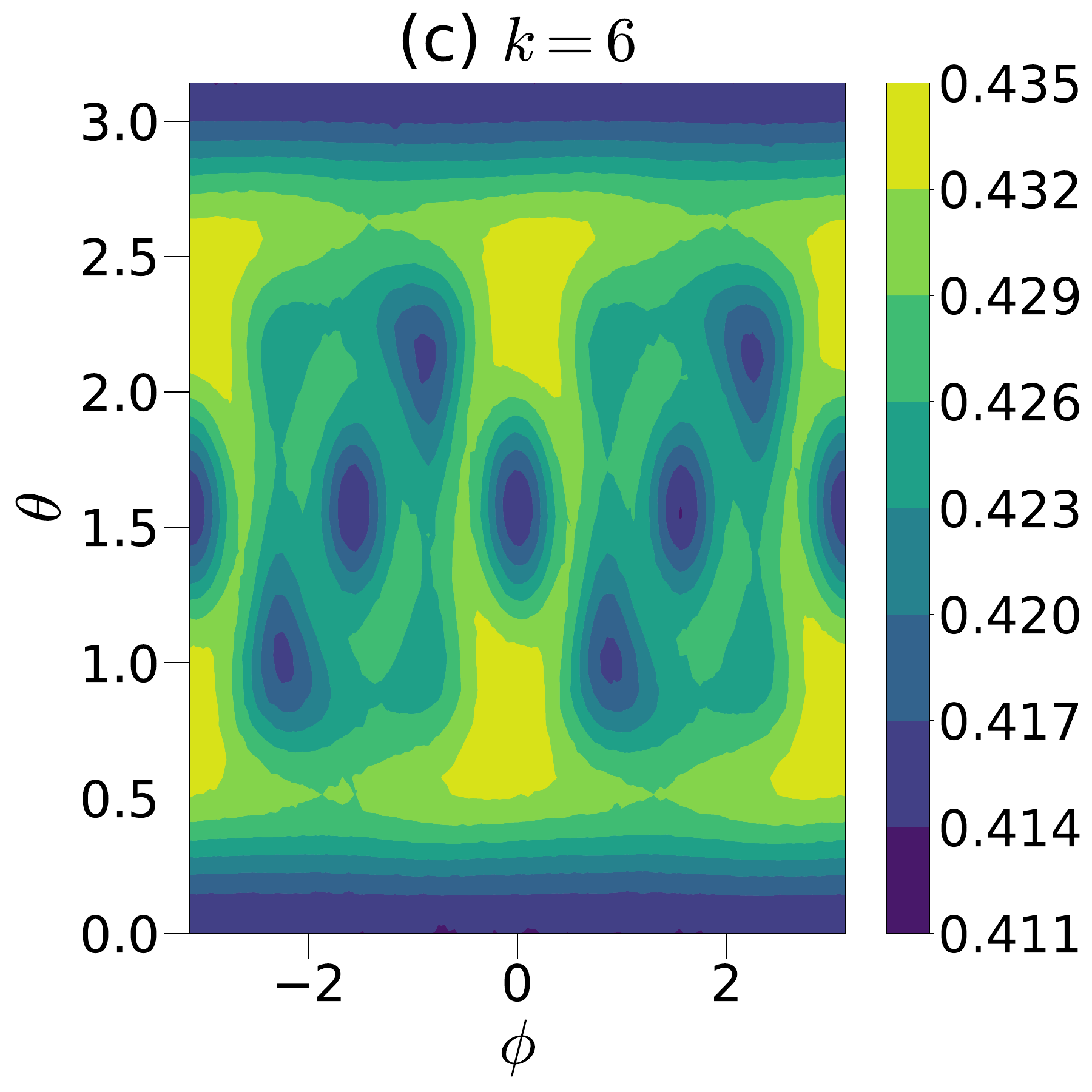}%
    \includegraphics[width=0.3\linewidth]{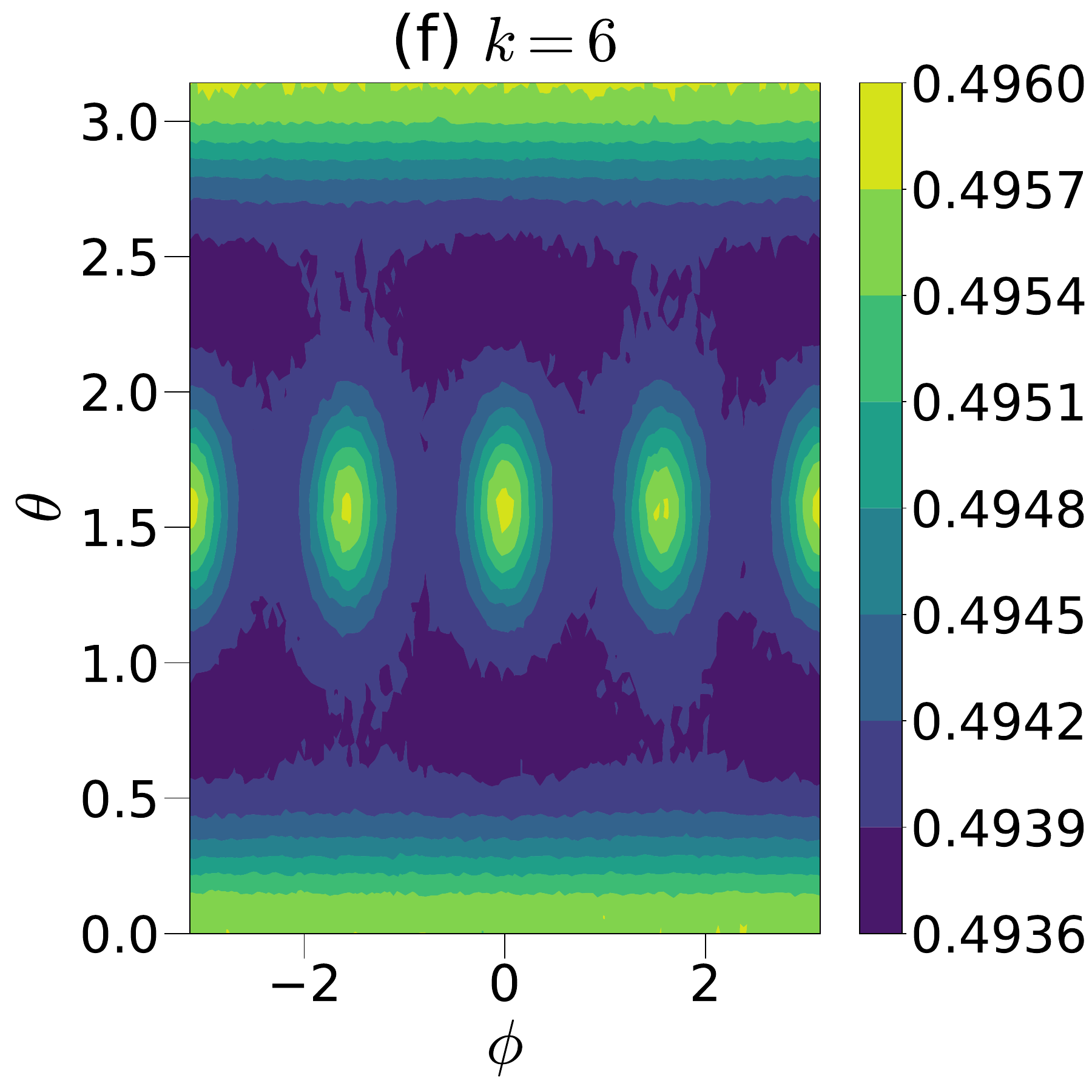}%
      \includegraphics[width=0.3\linewidth]{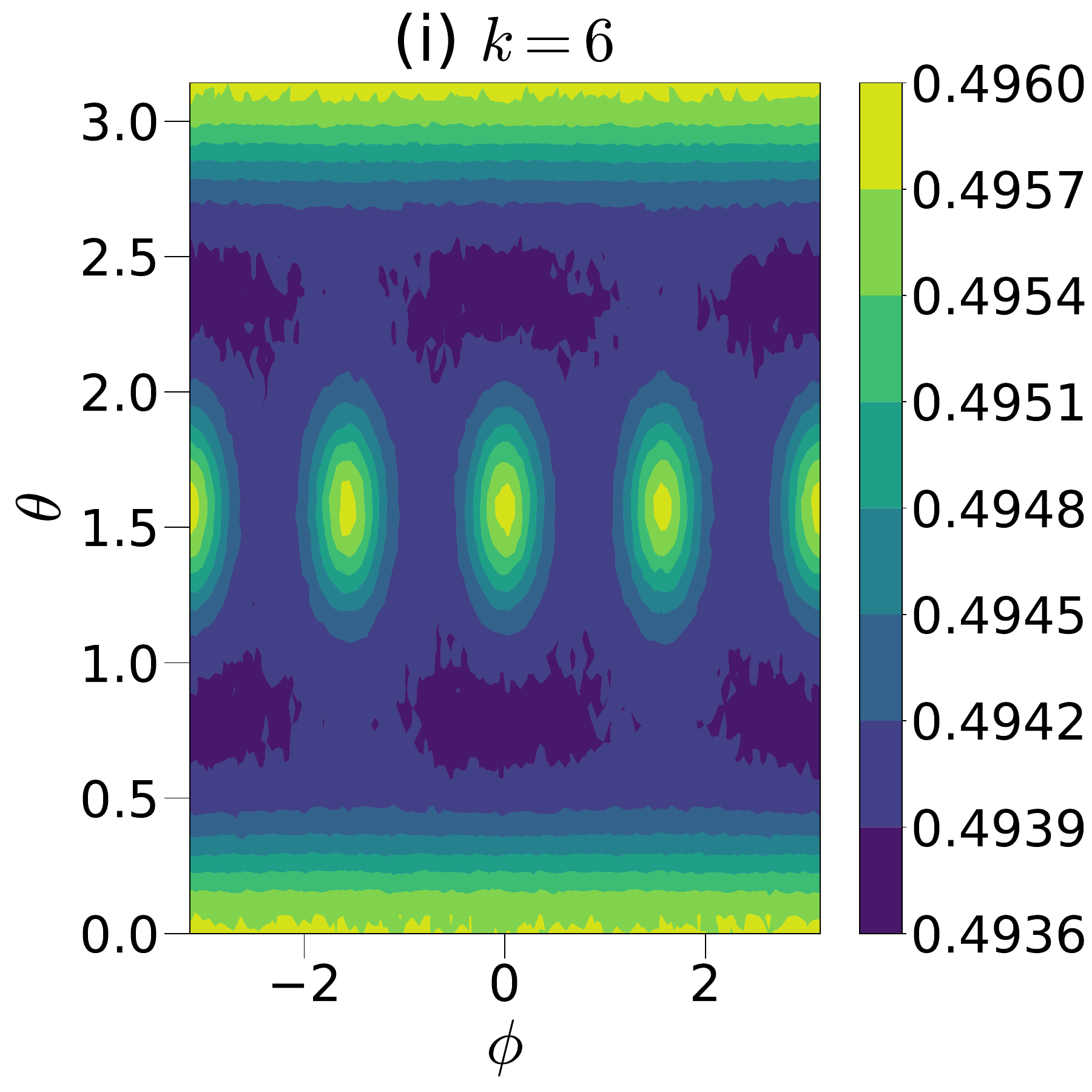}
        
    \caption{The plot shows the time and disorder averaged linear entropy for $k=1$, $k=3$ and $k=6$ for (column-wise starting from left) $w=0.01$, $w=1.0$ and $w=6.0$ with 5000 different initial conditions of $(\theta,\phi)$ from the interval $\theta$  $\in$  $(0,\pi)$ and $\phi$  $\in$  $(-\pi,\pi)$ for $N=8$. \textcolor{black}{Left most column belonging to $w=0.01$ is nearly indistinguishable to the disorder-free $w=0$ case, given in Fig.~4. Please note the different range of entropy values in each contour plot which varies with $k$ and $w$.}}
    \label{quantum_phase_space_1spin}
\end{figure*}

We now study the long-time averaged entanglement of the $Q:N-Q$ bipartition via the linear entropy $\langle \overline{S_Q} \rangle_w$  in the presence of disorder.  This quantity as a function of $w$ is shown in Fig.~\ref{long_time_disorder_int}(a) for $1:N-1$ partition and in Fig. \ref{long_time_disorder_int}(b) for $N/2:N/2$ partition.
We find that the averaged linear entropy for $k=6$ and $k=3$ monotonically grows with $w$ and saturates at large $w$. The saturation value corresponds to the linear entropy of $Q$ qubit subsystem in a random $N$ qubit state described by the Wishart ensemble of RMT in the full $2^{N}$ dimensions, and is given by \cite{page1993average, lubkin, tripartitescrambling, wishart1928generalised}
\begin{equation}\label{entropy}
\left \langle S_{Q} \right \rangle_{W,2^{Q}}= \frac{(2^{Q}-1)(2^{N-Q}-1)}{2^{N}+1},
\end{equation}
which is shown by the dashed horizontal line in Fig. \ref{long_time_disorder_int}. The value corresponding to the disorder free case $(w=0)$ in the chaotic limit using the permutation symmetric ensemble of $N+1$ dimension is given as \cite{tripartitescrambling}:
\begin{equation}\label{entropy_ps}
\left \langle S_{Q} \right \rangle_{PSS}= \frac{Q (N-Q)}{(Q+1)(N-Q+1)}
\end{equation}

Thus for $N=14$, $\left \langle S_{1} \right \rangle_{W,2} = 0.499$ and $\left \langle S_{7} \right \rangle_{W,2^{N/2}} = 0.984$, which are the observed saturation values for all $k$ for sufficiently large $w$. The corresponding values in PSS is given by $\left \langle S_{1} \right \rangle_{PSS} = 0.464$ and $\left \langle S_{7} \right \rangle_{PSS} = 0.766$.
 While there is a deviation from random state values, we find that the difference between numerical values and the random PSS value $\left \langle S_Q \right \rangle_{PSS}$ in Eq.~\ref{entropy_ps} decreases as $N$ or $k$ increases even more.
These results further confirm the observation that with increasing disorder strength, the system is eventually taken to the chaotic limit described by random matrix theory in full Hilbert space dimension independent of $k$. In other words, zero disorder, large $k$ values show average linear entropy given by the permutation symmetric ensemble of RMT whereas for large $w$, the linear entropy saturates to the Wishart ensemble of full $2^{N}$ dimensions of RMT.

The above calculations are performed for certain initial states as discussed in section \ref{longtimeaverage}. 
In order to get an idea of the evolution of other initial PSS states, we
compute the averaged linear entropy for all possible coherent states which forms a pseudo phase-space. The resultant structures can be 
compared with those of the disorder free case (Fig.~\ref{quantum_phase_space}), as well as with the classical phase space shown in Fig.~\ref{classical_phase_space}. Here also, we take the time average of linear entropy for a system of size $N=8$ upto 5000 kicks and 100 disorder averaging for 5000 initial values of $\theta,\phi$ from the interval $\theta$  $\in$  $(0,\pi)$ and $\phi$  $\in$  $(-\pi,\pi)$. The  ``phase-space" plots of $S_1$ for different $k$ and three values of disorder are shown in Fig.~\ref{quantum_phase_space_1spin}.

We find that the structures at $k=1$, $3$ and $6$ for lower disorder strengths with $w=0.01$ (left column) are %for smaller and larger variances, we find that for lower variances (left) the structures are.
similar to the ones in Fig.~\ref{quantum_phase_space} for the respective $k$ values without disorder, both in terms of magnitude of $S_1$ and features. %The lower bound of entanglement values has increased from 0 in $k=1$ case to 0.44 in $k=6$. 
For an intermediate value of disorder strength $w=1$ (middle column), the $k=1$ plot shows relatively no difference in the value of $S_1$ as well in the structure when compared to $w=0.01$. On the other hand, $k=3$ shows significant increase in the value of $S_1$ with the structures remaining similar to the $w=0.01$ case. Lastly, $k=6$ has completely changed at $w=1.0$, with higher values of entropy and completely different structures. 

For the largest disorder considered here, $w=6$ (rightmost column), we find all the three different $k-$values plots resemble with each other, though the figure for $k=1$ is less sharp. We have verified that $k=1,~w=6$ becomes more sharp when $w$ is further increased.
Clearly,  there is no difference between $k=6,~  w=1$ and $k=6,~w=6$, which seems to be the final possible phase space structure.
%Let us now discuss the values in the large disorder limit (rightmost column). 
The RMT value corresponding to full Hilbert space is equal to $\approx 0.494$. While most of the initial states do saturate to this value (blue colored region), we find small deviations around fixed points and points corresponding to period-4 orbit. These periodic orbits are associated with the symmetries of the Floquet operator $U_0$ for the disorder free case $w=0$ as discussed in \cite{haake1987classical}. However these symmetries are also preserved in the disordered case as well. Hence these points show distinct behavior in the large disorder case too, and indicate the robustness of these structures in the presence of disorder.

\section{ The case of $p=4\pi/11$}
\label{4pi/11}

 We now consider a rotation angle around the $y-$ axis that is slightly different from $\pi/2$. This is for two reasons:
the first is to ensure the robustness of our studies for different values of the rotation angle $p$. The second is that the reduced symmetries allow for a more easy comparison with standard random matrix spectral statistics, and not only the state properties discussed above.

\begin{figure}[H]
\centering
\includegraphics[scale=0.2]{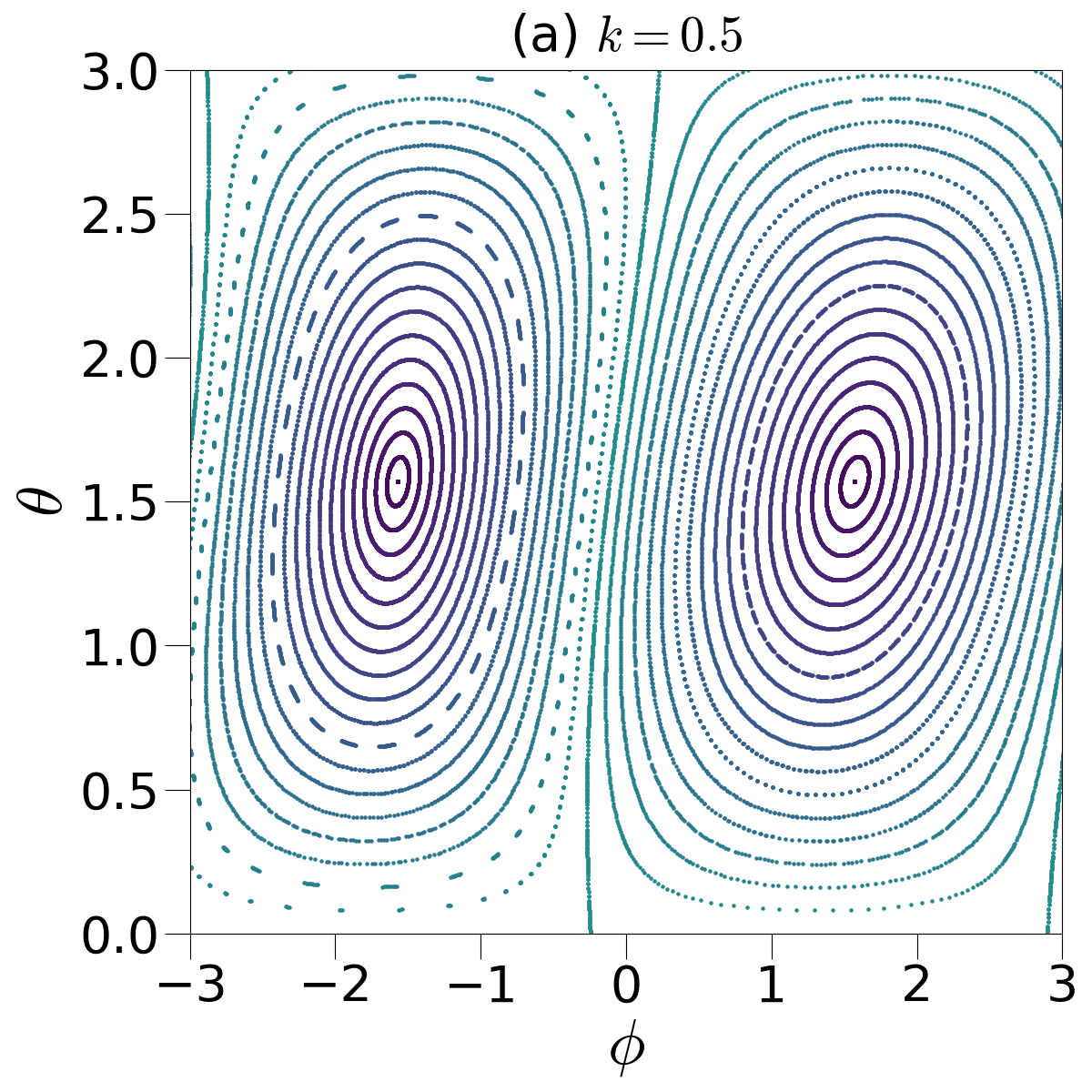}
\includegraphics[scale=0.2]{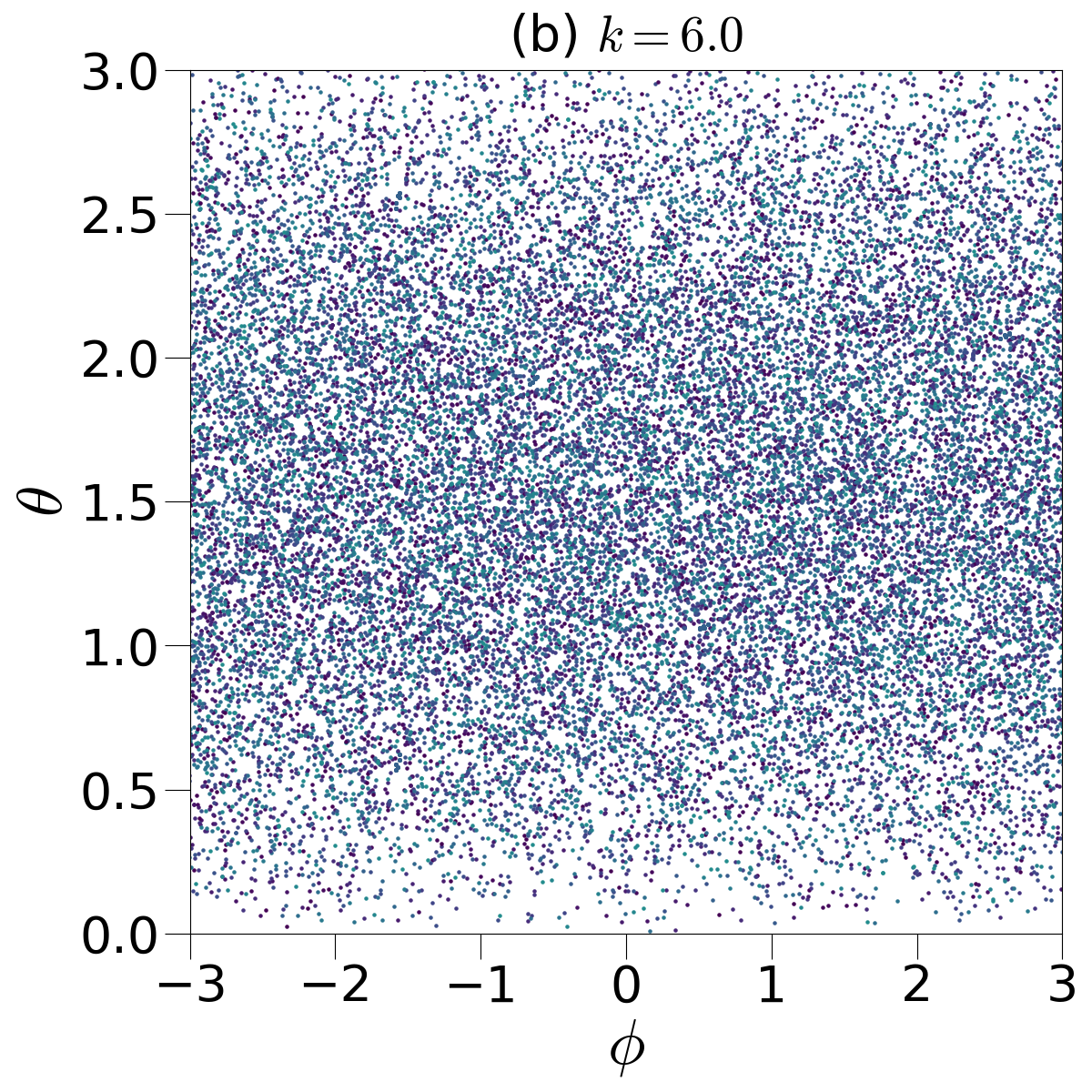}
  \caption{Classical phase space for $p=4\pi/11$, (a). $k=0.5$ that shows predominantly regular behavior, (b). $k=6$ where it becomes fully chaotic.}
  \label{cp_diffp}
\end{figure}

We focus only on two values of $k$, one corresponding to the regular dynamics ($k=0.5$) and the other corresponding to the chaotic dynamics $(k=6.0)$.
The disorder free classical phase space plots for this case is given in Fig.~\ref{cp_diffp}. We see that for low chaos parameter $k=0.5$, the phase space is predominantly regular and for $k=6$, it becomes fully chaotic. In this case also, we find the fixed points to be $(\pi/2,\pi/2)$ and $(\pi/2,-\pi/2)$ that are stable at low $k$ values.  In contrast to the $p=\pi/2$ case, there are no period-4 orbits at the $x-$ and $z$ poles. 

\begin{figure}[ht]
\centering
   \includegraphics[width=1.0\linewidth, height=0.8\linewidth]{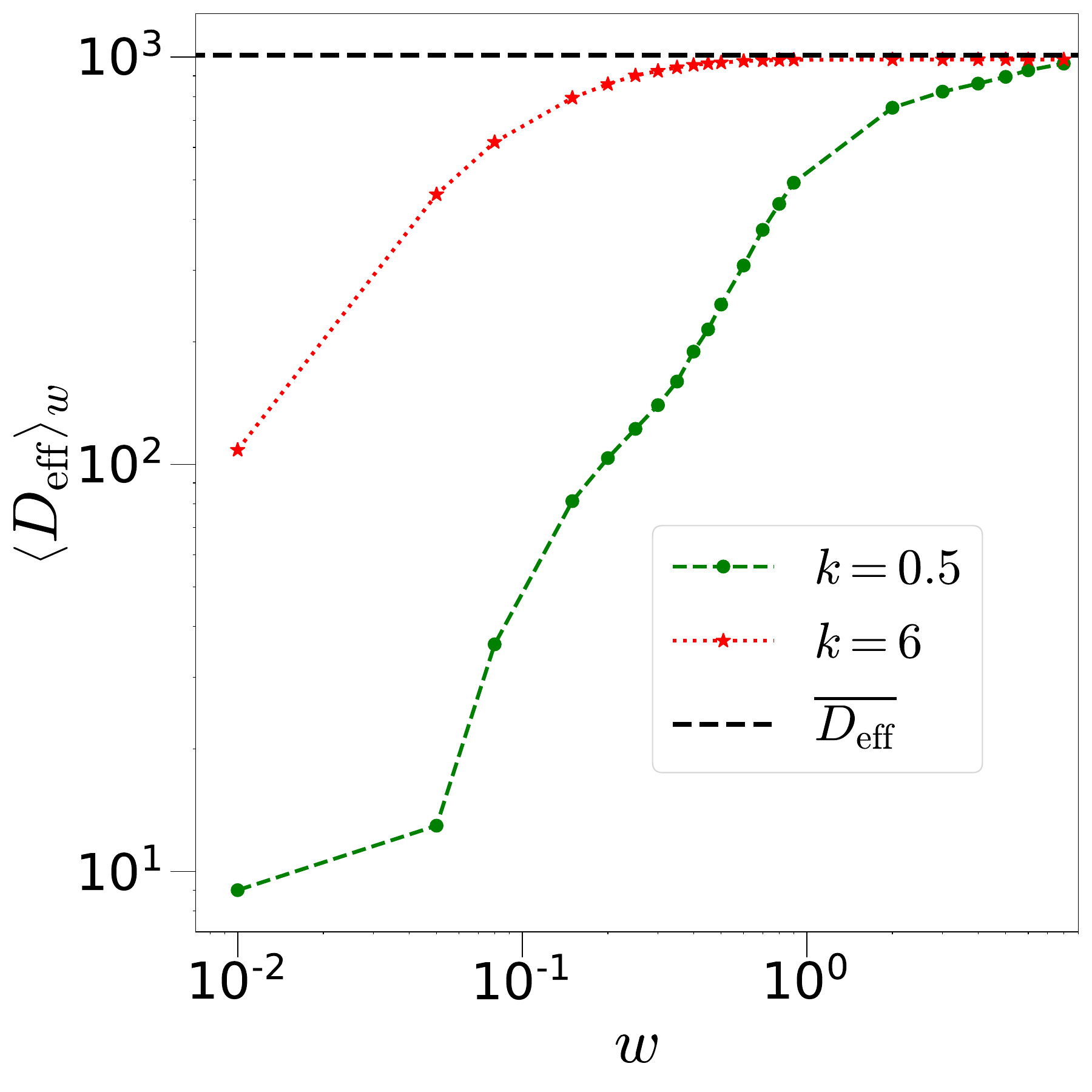}
  \caption{$\langle D_{\rm eff} \rangle_w$ for $k=0.5$ and $k=6$ \textcolor{black}{increases towards $\overline{D_{\rm eff}}$} as the strength of disorder $w$ is increased. The system size is $N=10$ qubits and the initial state is $\ket{\theta,\phi}=\ket{2.25,1.1}$.  }
  \label{eff_dim_diffp}
\end{figure}

 Figure~\ref{eff_dim_diffp}.
shows the effective dimension  $D_{\rm eff}$ of coherent states in eigenstate basis 
for the disordered quantum cases with $k=0.5$ and $k=6$. Similar  results to that of $p=\pi/2$ case is observed: an increase with disorder reaching \textcolor{black}{nearly} the full Hilbert space dimension $\sim 2^N$. Compared to low values of $k$, a relatively weaker disorder strength can drive the dynamics across the entire Hilbert space for larger $k$ values.

\begin{figure}[h]
\centering
    \includegraphics[width=1.0\linewidth, height=0.8\linewidth]{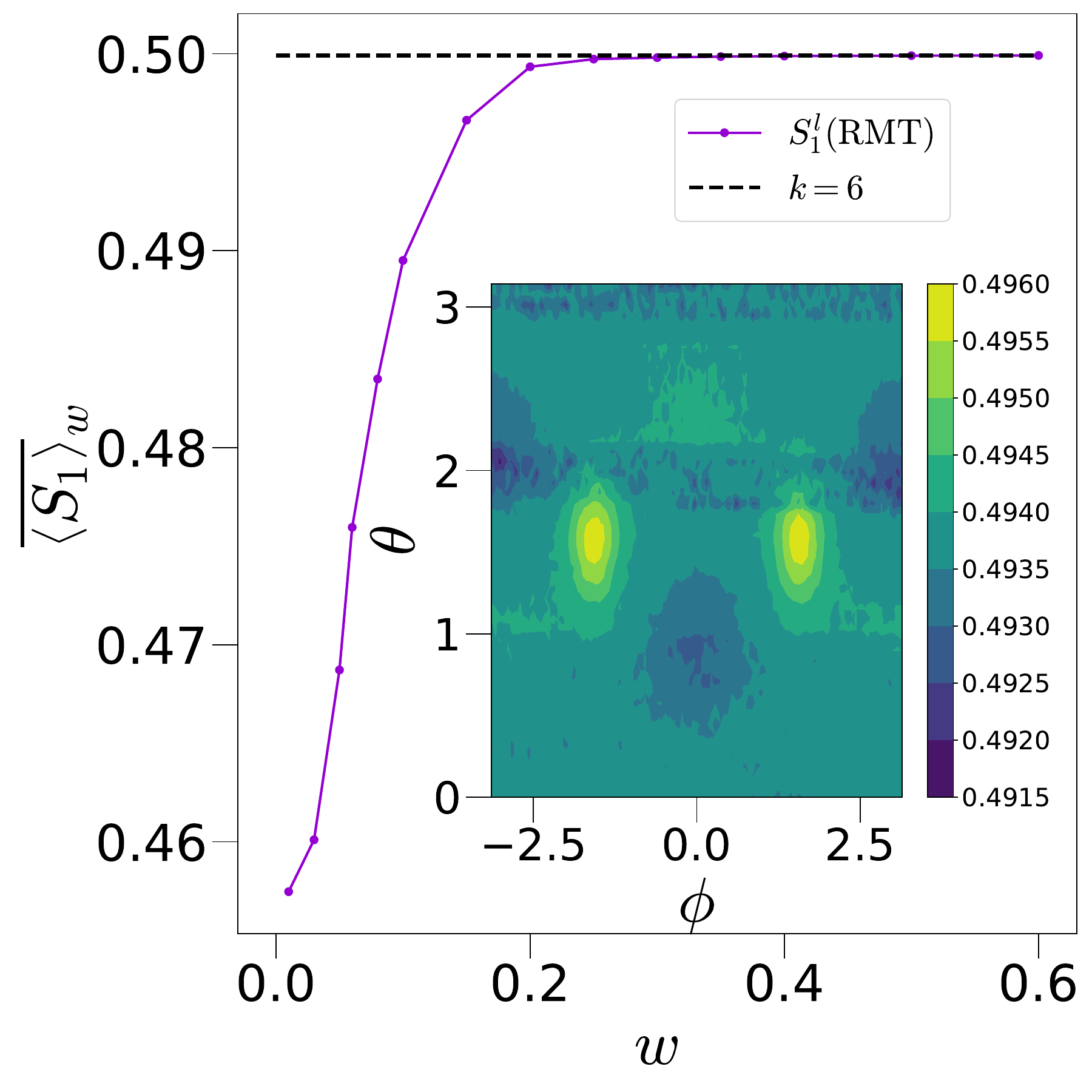}
   \caption{The time and disorder averaged linear entropy in $1:N-1$ bipartition with respect to disorder $w$ for $N=14$, $k=6$ and initial state $\ket{2.25,1.1}$. Inset shows the same for 5000 different initial conditions of  $(\theta,\phi)$ from the interval $\theta$  $\in$  $(0,\pi)$ and $\phi$  $\in$  $(-\pi,\pi)$ for $N=8$, $k=6$ and $w=6$.}

  \label{phase_space_diffp}
\end{figure}

Figure~\ref{phase_space_diffp} shows the time and disorder averaged linear entropy at $k=6$, when the initial state is $\ket{2.25,1.1}$. 
%We find that 
Similar to the $p=\pi/2$ case, as the disorder is increased, the entropy saturates to the RMT value of the full Hilbert space dimension. 
%Similar to the $p=\pi/2$ case, 
%at the phase space 
We also plot the time averaged linear entropy for different initial coherent states at $k=6$ and at large disorder $w=6$ for $N=8$ in the inset of Fig.~\ref{phase_space_diffp}. Here we see that the states corresponding to  $(\pi/2,\pi/2)$ and $(\pi/2,-\pi/2)$ have slightly higher entropy than the RMT value. These are the fixed points in the classical phase space at small $k$ values which loses stability and vanishes in the large $k$ limit. However their  remnants are still visible in the  average linear entropy with enhanced entanglement values pertaining to these initial conditions. Notable is the absence of the structures at the $z-$ and $x-$ poles observed for the $p=\pi/2$ case as shown in Fig.~\ref{quantum_phase_space_1spin}(i).

\subsection{Spectral statistics}

\begin{figure}[h]
\centering
  \includegraphics[width=1.0\linewidth, height=0.8\linewidth]{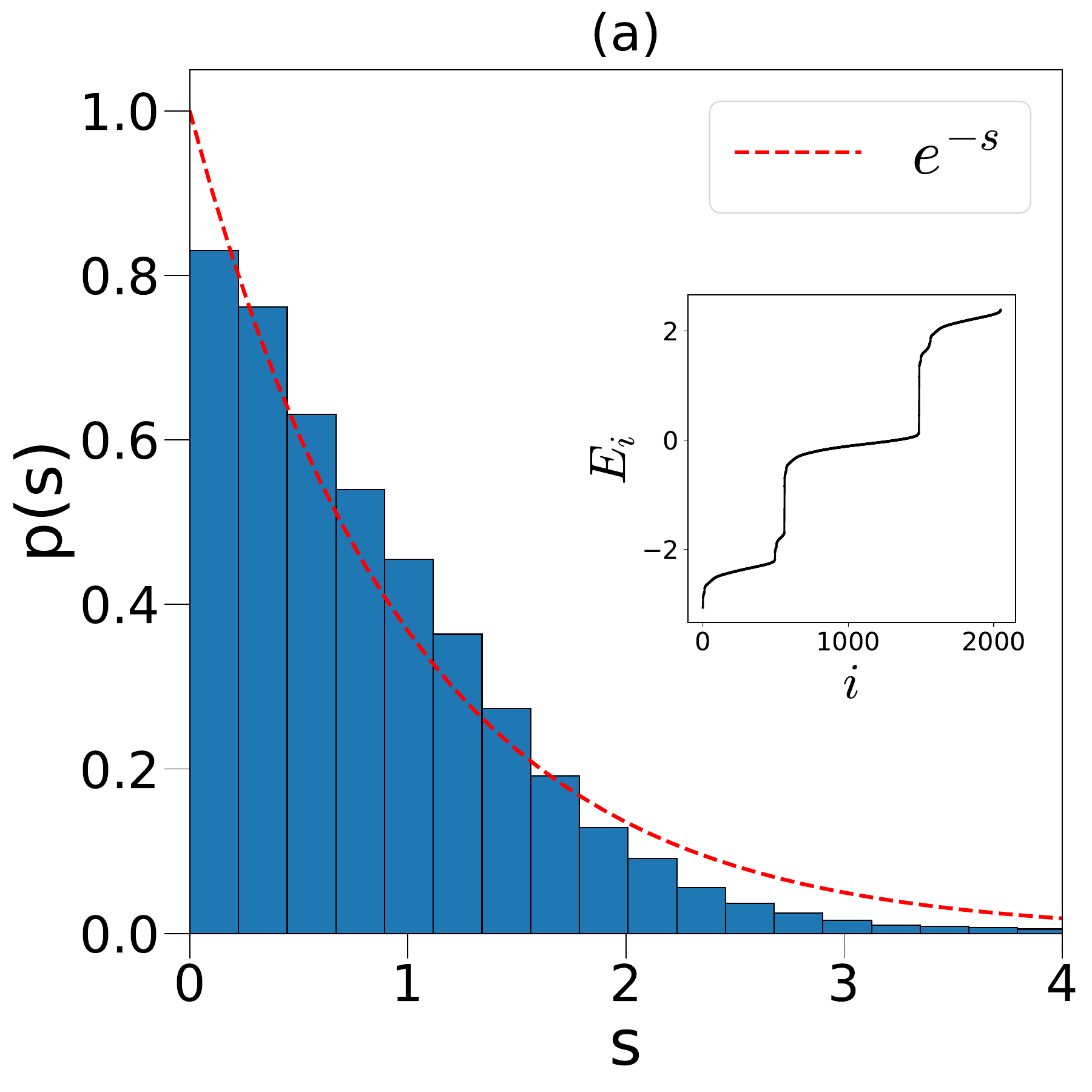}
  \includegraphics[width=1.0\linewidth, height=0.8\linewidth]{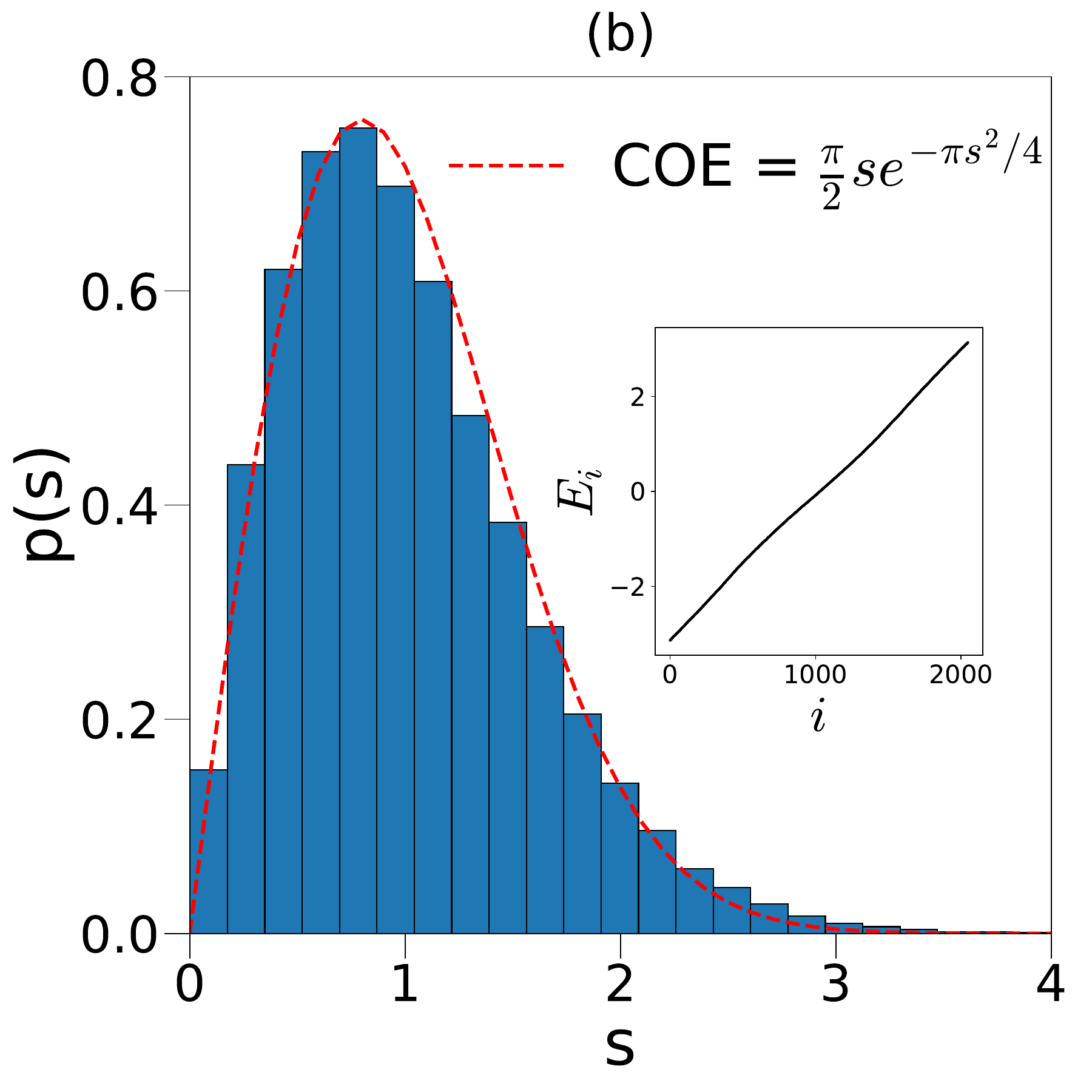}
   \caption{The spacing statistics for $N=12$ at $k=0.5$ for (a) small disorder $w=0.5$ and (b) large disorder $w=8.0$ for 50 disorder realisations. The dashed lines shows the spacing statistics for the Poisson and Wigner Dyson (COE) distribution. The inset shows the density profile for one disorder realization in each case.}

  \label{statistics_diffp}
\end{figure}

We now analyze the spectral statistics \cite{Haakebook, mehta2004random} in order to further confirm the transition to a chaotic phase with increase in disorder $w$.  
It is well established that the level spacing statistics for disorder free case $w=0$ at $k=0.5$ exhibits Poisson distribution \cite{wang2023statistics}.
Here, the Floquet operator $U_0$ commutes with the parity operator $\hat{R}=e^{-i\pi J_y}$, thus the matrix can be block diagonalised into even and odd parity blocks.
In the disordered case also, parity symmetry holds and we use the positive parity sector for the spectral analysis.   For small values of the disorder strength $w$ the $2^N$ eigenangles have high degeneracy as the permutation symmetry is not fully broken. For intermediate $w$ there is an approximate band structure corresponding to the irreducible representations of spin $S$ with $S=0, \cdots, N/2$. As the disorder increases, the bands merge and become the uniform density that one expects in Floquet systems. Due to the atypical nature of this transition that includes breaking permutation symmetry, the spectral statistics can show unusual features. We present two cases in Fig. \ref{statistics_diffp} where $w=0.5$ exhibits near Poisson level statistics (after unfolding), whereas in the limit of large disorder it shows Wigner-Dyson distribution corresponding to the circular orthogonal ensemble (COE). The insets in these figures show the spectrum of quasienergies or eigenangles. The unusual energy distribution with distinct clusters when $w=0.5$ is a consequence of the symmetry that is not fully broken. Thus these preliminary findings indicate a rich phenomenology in the spectral statistics of these systems as well.

\section{Disorder in field: $p=\pi/2$}
\label{sec_disorder_field}

\begin{figure}[h]
    \includegraphics[width=1.0\linewidth, height=0.8\linewidth]{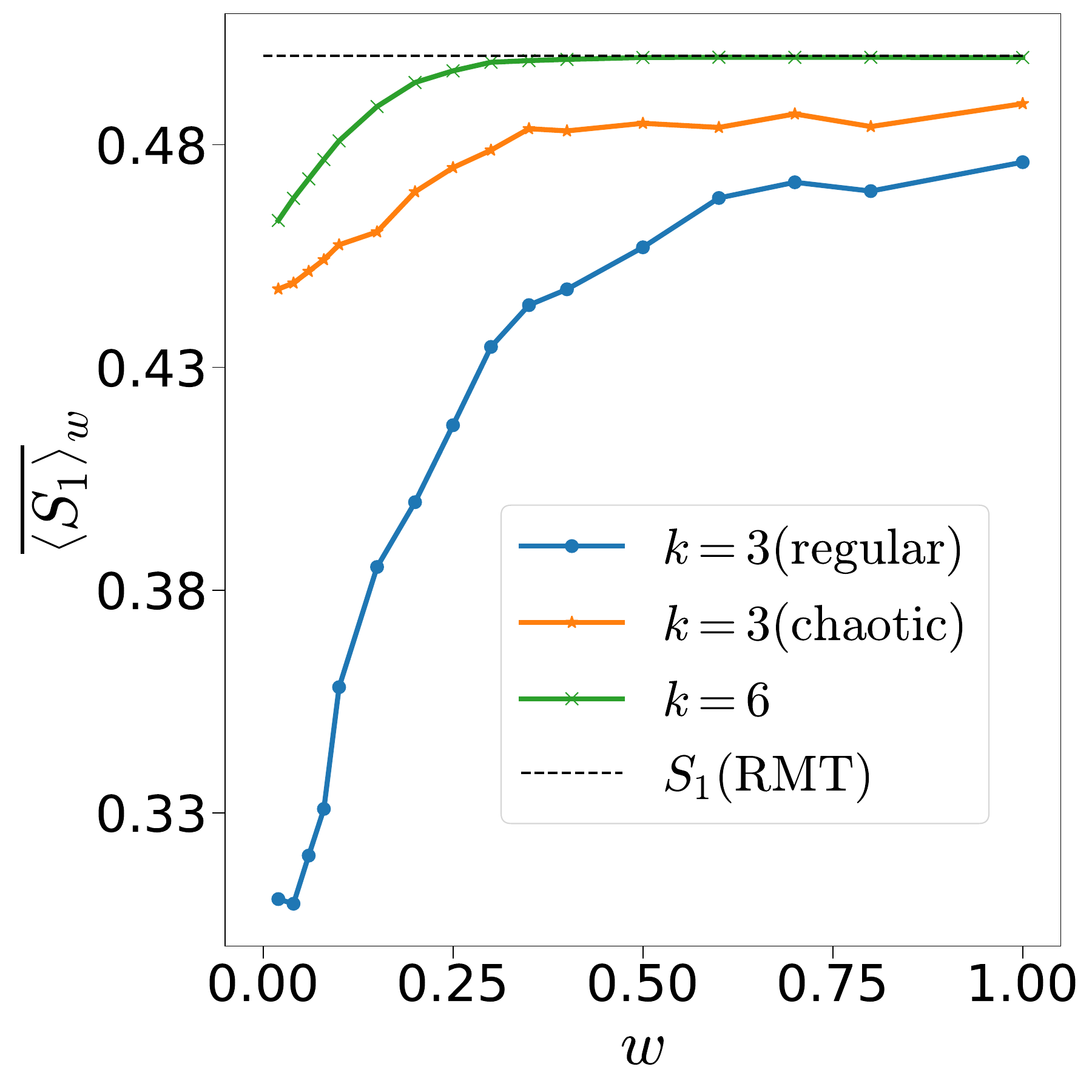}
    \caption{The long time and disorder averaged linear entropy in $1:N-1$ partition with respect to disorder strength $w$ for $N=14$ qubits at different $k$ values when the disorder is in the field term. We have checked that $k=3$ approaches the RMT values as $w$ is increased. }
\label{entropy_field_kicks}
\end{figure}

We now very briefly discuss the effect of disorder in the field term of the Hamiltonian, rather than in the interaction.
The Floquet operator 
is given by:
\begin{equation}
U'_{field}=\exp\left ({-i\frac{k}{2N}\sum_{l< l'=1}^{N}\sigma_{l}^{x}\sigma_{l'}^{x} }\right) \exp\left ({-i\frac{\pi}{4}\sum_{l=1}^{N}(1+\epsilon_{l})\sigma_{l}^{y}}\right ),
\end{equation}
where $\epsilon_{l}$ is a random number taken from a normal distribution with zero mean and standard deviation $w$.
%\subsection{Long time dynamics}
We focus only on the variation of long time averaged linear entropy as a function of $w$, which is plotted in Fig. \ref{entropy_field_kicks}. Here also, we see that the long time averaged linear entropy appears to saturate to Wishart ensemble of RMT in full $2^{N}$ dimension, albeit for larger $w$ values  (see Fig.~\ref{long_time_disorder_int}(a) for comparison). We expect almost similar behaviour in case of other measures when the disorder is present in the field term. 
 
\section{Summary and Discussions}In this work, we have studied the effect of breaking the permutation symmetry of the well-studied kicked top model.
As the model can be considered to be a long range interacting kicked spin system, this  symmetry breaking can be achieved by introducing disorder via interactions. 
The interplay between dynamical chaos and disorder related physics leads to a rich set of phenomena that this work has begun to study. \textcolor{black}{The kicked top is a generic nonintegrable Floquet system with an integrable to chaotic transition. Thus we expect our results to be applicable to a wide variety of Floquet permutation symmetric systems.}

 Introducing disorder in the system has two important effects: Firstly, the dynamics is taken out of the permutation symmetric subspace. This is shown using the overlap of time evolving states  with the permutation symmetric subspace ($\left \langle \chi \right \rangle_w$) as well as an effective dimension $\langle D_{\text{eff}} \rangle_w$ of coherent states in the eigenbasis. Secondly, the dynamics become more and more chaotic as $w$ is increased, even if the initial dynamical chaos present was small or even absent. Thus disorder becomes a proxy for chaos.
%till the long time entanglement measures saturate to a value given by the Wishart ensemble of RMT in full $2^{N}$ dimensions. 
This is confirmed in multiple interesting ways. For example, the gradual  destruction of quantum revivals that are present for regular dynamics as the disorder is increased. For small disorder, we continue to observe the revivals at a time scale $\tau_{\hbar} \sim {N}$. The shift to chaotic dynamics as disorder is increased, is also confirmed by spectral statistics which changes from Poisson at $w=0$ to Wigner Dyson corresponding to COE at large $w$.

\textcolor{black}{The other quantity that we study is entanglement in the form of linear entropy though we have verified that other entropy measures like the von Neumann entropy exhibits similar qualitative behaviour \cite{manju24}}.
While entanglement has long been studied in the quantum kicked top for the disorder free case, we have 
built quantitatively on previously known connections between entanglement and a classical diffusion of densities in phase space. Thus, we give analytical expressions for the one-qubit entanglement (via the linear entropy) for both the clean near integrable and chaotic cases. The effects of disorder on the entanglement growth is then studied both analytically and numerically. Analytically, we model disorder as resulting from noisy channel which gives simple expressions for entanglement growth involving diffusion constant. Numerically, the saturation value of linear entropy confirms that the nonequilibrium state at large disorder corresponds to random states on the full Hilbert space. We find a good agreement between the form of analytical expression and the numerical calculations. Nevertheless there are important differences in the secular growth phase with and without disorder.
We find a quadratic initial growth for a disorder free system which changes to linear for large disorder.
Thus the growth of entanglement for a clean chaotic kicked top is quite different from that of a disordered regular kicked top, even though they both happen in a short time, that is the Ehrenfest time for the clean case. 

We also plot pseudo-phase space structures using one qubit linear entropy for different initial coherent states, and show that in the limit of large disorder, most of them saturate to linear entropy corresponding to RMT in full Hilbert space. However there are scarring by fixed points even in the large disorder limit, which indicates another feature that deserves further study: the presence of quantum scars in these disordered many-body systems.
Experimental realisation of disorder free quantum kicked top has been done in Refs. \cite{chaudhury_nature_husimi, 2016_ergodic, NMRstudiesin2qubit}. We believe that such an experiment can be extended to the disordered case so that our results may be verified.
\label{conclusion}

\begin{acknowledgments}
MC and UD acknowledge the HPC facility Chandra at IIT Palakkad where the computations were carried out. UD and AL acknowledge discussions during Quantum Many-Body Physics in the Age of Quantum Information (code: ICTS/qmbpqi2024/11) organized by International Centre for Theoretical Studies, India. 
\end{acknowledgments}
%\clearpage
\appendix
\section{Disorder free case: Regular regime}
Let us assume an initial density that is a Gaussian distribution in action-angle variables:
\begin{equation}
\rho(\theta,I,0)=\frac{1}{2 \pi \sigma^2 } \sum_{l=-\infty}^{\infty}e^{-(\theta +2 \pi l)^2/(2 \sigma^2)}e^{-(I-I_0)^2/(2 \sigma^2)}. 
\end{equation}
Assuming that the action-angle variables are ``good" in the near integrable regimes, we have from Liouville theorem that 
$ \rho(\theta,I,t)= \rho(\theta-\omega(I) t,I,0)$.

As the nontrivial evolution is only in angle, the density diffuses on a tight bundle of tori. Taking a classical observable $f(\theta)$, its expectation value can be calculated as
\begin{eqnarray}
\langle f \rangle(t) &= \frac{1}{2 \pi}
 \int_{-\pi}^{\pi} \int_{-\infty}^{\infty}f(\theta) \rho(\theta,I,t) d\theta dI.
\end{eqnarray}
Expanding in a Fourier series $f(\theta)=\sum_{k=-\infty}^{+\infty} f_{k} e^{-ik\theta} $,
we get $\langle f \rangle(t)$
%\begin{align}
%\langle f \rangle(t) =\int_{-\pi}^{\pi}\int_{\infty}^{\infty}\sum_{k=-\infty}^{+\infty}f_{k}e^{-ik\theta} \rho(\theta,I,t) d\theta dI \nonumber.
%\end{align}

\begin{equation}
 =\int_{-\pi}^{\pi}\int_{\infty}^{\infty}\sum_{k=-\infty}^{+\infty}f_{k}e^{-ik\theta} \rho(\theta-\omega(I) t,I,0) d\theta dI
\end{equation}

With $\theta'=\theta-\omega(I) t  $
and approximating $\omega(I) \approx \omega(I_0)+\omega'(I_0)(I-I_0)$, $\langle f \rangle(t)$ takes a form
%\[ =\sum_{k=-\infty}^{+\infty}\int_{-\pi}^{\pi}\int_{\infty}^{\infty}f_{k}e^{-ik(\theta'+\omega(I) t)} \rho(\theta',I,0) d\theta' dI\]

\begin{align}
=&\sum_{k=-\infty}^{+\infty}\int \int f_{k}e^{-ik(\theta'+\omega(I_0)t+\omega'(I_0)(I-I_0) t)} \rho(\theta',I,0) d\theta' dI \nonumber \\
=&\frac{1}{2\pi \sigma^2}\sum_{k} f_ke^{-ik\omega(I_0)t} \sum_{l}\int_{-\pi}^{\pi} e^{-ik\theta't} e^{-\frac{(\theta'+2\pi l)^2}{2\sigma^2}} d\theta' \nonumber \\
& \times \int_{-\infty}^{\infty}e^{-ik\omega'(I_0)(I-I_0)t} e^{-\frac{(I-I_0)^2}{2\sigma^2}}dI.
\end{align}
Using 
$\int_{-\infty}^{\infty} e^{-ikx}e^{-\frac{(x+a)^2}{b^2}}dx=b\sqrt{\pi}e^{iak}e^{\frac{k^2b^2}{4}}$,\\\\
we finally get 
\begin{equation}
\left \langle f \right \rangle (t)=\sum_{k}f_{k} e^{-ik\omega(I_0)t} e^{-\frac{k^2\sigma^2}{2}} e^{-\frac{k^2 \omega'(I_0)t^2 \sigma^2}{2}}.
\end{equation}
\label{appendix_A}

%\appendix
\section{Disorder free case: Chaotic regime}
For the chaotic regime in the disorder free limit, we evaluate the variance for a specific solvable model which shows complete chaos, and argue why we should expect this to be the generic behavior.
The cat maps are automorphisms of the unit torus and given by 
\begin{equation}
\begin{pmatrix}
q_{n+1} \\
p_{n+1} 
\end{pmatrix}
=
\begin{pmatrix}  
a & b \\
c & d
\end{pmatrix}
\begin{pmatrix}
q_n \\
p_n 
\end{pmatrix} \mod 1.
\end{equation}
where $a,b,c,d$ are integers such that $ad-bc=1$. If $|a+d|>2$ the map is fully chaotic, for example the Arnold cat map has $(a=2,\, b=1,\, c=1,\,d=1)$ \cite{cat_map}. The general case is solved by 
\begin{equation}
\begin{split}
\begin{pmatrix}
q_{n} \\
p_{n} 
\end{pmatrix}
&=
\begin{pmatrix}  
a & b \\
c & d
\end{pmatrix}^n
\begin{pmatrix}
q_0 \\
p_0 
\end{pmatrix} \mod 1\\
& \equiv \begin{pmatrix}  
a_n & b_n \\
c_n & d_n
\end{pmatrix}
\begin{pmatrix}
q_0 \\
p_0 
\end{pmatrix} \mod 1.
\end{split}
\end{equation}
If the map is chaotic, each of the matrix elements $a_n, b_n$ etc. increase exponentially with time $n$. In the Arnold cat map case, $a_n=F_{2n+1}$, $b_n=c_n=F_{2n}$ and $d_n=F_{2n-1}$, where $F_k$ is the $k^{\mbox{th}}$ Pingala-Fibonacci number with $F_1=1, F_2=1$ and $F_{n+1}=F_{n}+F_{n-1}$.

Let the initial density be 
%\begin{equation}
%\rho(q,p)=\frac{1}{2 \pi \sigma^2 }\sum_{m,n} e^{-(q+m-q_0)^2/(2 \sigma^2)}e^{-(p+n-p_0)^2/(2 \sigma^2)}
%\end{equation}
%Also, 
\begin{equation}
\rho(q,p,0)=\sum_{k,l=-\infty}^{\infty}a_{kl}(0) e^{2\pi i (qk+pl)},
\end{equation}
where 
\begin{equation}
\begin{split}
a_{kl}(0)&=\int \int \rho(q,p,0)e^{-2\pi i (qk+pl)} dqdp  \\
&=e^{-2\pi i (q_0k+p_0l)}e^{-2\pi \sigma^2(k^2+l^2)}.
\end{split}
\end{equation}
This ensures that the initial density is a periodicised Gaussian with width $\sigma$ and centered at $(q_0,\,p_0)$. From Liouville theorem it follows that the density at time $n$ is 
\begin{equation}
    \rho(q,p,n)=\sum_{k,l=-\infty}^{\infty}a_{kl}(n) e^{2\pi i (qk+pl)},
\end{equation}
with $a_{kl}(n)=a_{k_n l_n}(0),$ where
\begin{equation}
\begin{pmatrix}
k_n \\
l_n 
\end{pmatrix}=
\begin{pmatrix}  
a_n & b_n \\
c_n & d_n
\end{pmatrix}
\begin{pmatrix}
k \\
l 
\end{pmatrix}.
\end{equation}
This shows that except for the $a_{00}(0)=a_{00}(n)$ mode, the others develop exponentially large frequencies in phase space and the density is to a good approximation coarse grained to the uniform distribution.

If $f(q,p)=\sum_{kl}f_{kl} e^{2\pi i (kq+lp)}$ be any observable, its variance 
is then evaluated as
\begin{equation}
    \mbox{var} \, f(n)= \sum_{k,l,k',l'} f_{kl}f_{k'l'}\left[ a_{k+k'\, l+l'}(n)-a_{kl}(n)a_{k'l'}(n) \right]
\end{equation}
For simplicity considering the case $f(q,p)=\cos(2\pi q)$ and $p_0=0$,  we get 
\begin{equation}
\begin{split}
    \mbox{var} \, f(n)=&\frac{1}{2}\left(1-e^{-4 \pi^2 \sigma^2(a_n^2+c_n^2)}\right)\\
    \times & \left(1- \cos(4 \pi q_0 a_n) e^{-4 \pi^2 \sigma^2(a_n^2+c_n^2)}\right)
    \end{split}
\end{equation}
If $a_n^2\sim e^{2\lambda n}$, where $\lambda$ is asymptotically the Lyapunov exponent we get a rapid equilibration of the variance to $1/2$. 

While we have derived this for the cat map, it may be argued that for generic chaotic systems, Gaussian densities that have small widths will have a linear evolution around the trajectory of $(q_0,p_0)$, and when nonlinear effects are seen are the effective time at which the widths are of order 1, when the ``log-time" as discussed in the text is reached. This motivates the form assumed for the variance and hence the entanglement growth in the main text. 
\label{appB}
%\appendix
\section{Disordered case}
\label{AppC}
In order to understand the effect of disorder in the integrable dynamics, we mimic the disorder by introducing noise in the system. If the rotation map is given in action-angle variables as $\theta_{n+1}=\theta_n+\omega +\xi_n$, where $\xi_n$ is taken as an i.i.d. random variable, we can write after a large time $t$, $\theta(t)$ as
\begin{equation}
\theta(t)=\theta(0)+\omega(I)t+\eta(t)
\end{equation}
where $\eta(t)$ is the cumulative noise with a probability distribution \[ P(\eta)=\frac{1}{\sqrt{2\pi Dt}}e^{-\frac{\eta^{2}}{2Dt}}.\]
Here $D$ is a diffusion coefficient.

\begin{equation}
\begin{split}
\left \langle f\right \rangle(t) &= \frac{1}{2\pi}\int_{-\pi}^{\pi}f(\theta)\rho(\theta,I,t)\, d\theta \, dI\,  P(\eta) d\eta 
%\\
%&=\frac{1}{2\pi}\int_{-\infty}^{\infty}\int_{-\infty}^{\infty}\int_{-\pi}^{\pi}\sum_{k}f_ke^{-ik\theta}\rho(\theta(0),I,0)d\theta dI d\eta\\
%&=\frac{1}{2\pi}\int\sum_{k}f_ke^{-ik\theta}\rho(\theta-\omega(I)t-\eta(t),I,0)d\theta dI d\eta 
\end{split}
\end{equation}
%Let $\theta(t)-\omega(I)t-\eta(t)=\theta'$
Following the steps outlined in Appendix \ref{appendix_A} and integrating over the noise results in 

\begin{align}
%&=\frac{1}{2\pi}\int \int \int \sum_{k}f_ke^{-ik(\theta'+\omega(I)t+\eta(t))}\rho(\theta',I,0)d\theta' dI d\eta \nonumber \\
%&=\sum_{k}f_k e^{-ik\omega(I_0)t} e^{-\frac{k^2 \sigma^2}{2}(1+\omega'(I_0)^2t^2)}  \int \frac{1}{\sqrt{2\pi Dt}}e^{-ik \eta(t)}e^{-\frac{\eta^2}{2Dt}} d\eta  \nonumber \\
\left \langle f\right \rangle(t)&=\sum_{k=-\infty}^{\infty}f_k e^{-ik\omega(I_0)t} e^{-\frac{k^2 \sigma^2}{2}(1+\omega'(I_0)^2t^2)} e^{-\frac{k^2Dt}{2}}.
\end{align}
For the specific case of $f(\theta)=\cos(\theta)$, we get:
\begin{equation}
\begin{split}
&\mbox{var}\,f(t)=\frac{1}{2}(1-e^{-Dt}e^{-\sigma^2(1+\omega'(I_0)^2t^2)}) \\ & \times (1-e^{-Dt}e^{\sigma^2(1+\omega'(I_0)^2t^2)}\cos(2\omega(I_0)t)).
\end{split}
\end{equation}

\begin{figure}
 \includegraphics[width=1.0\linewidth, height=0.8\linewidth]{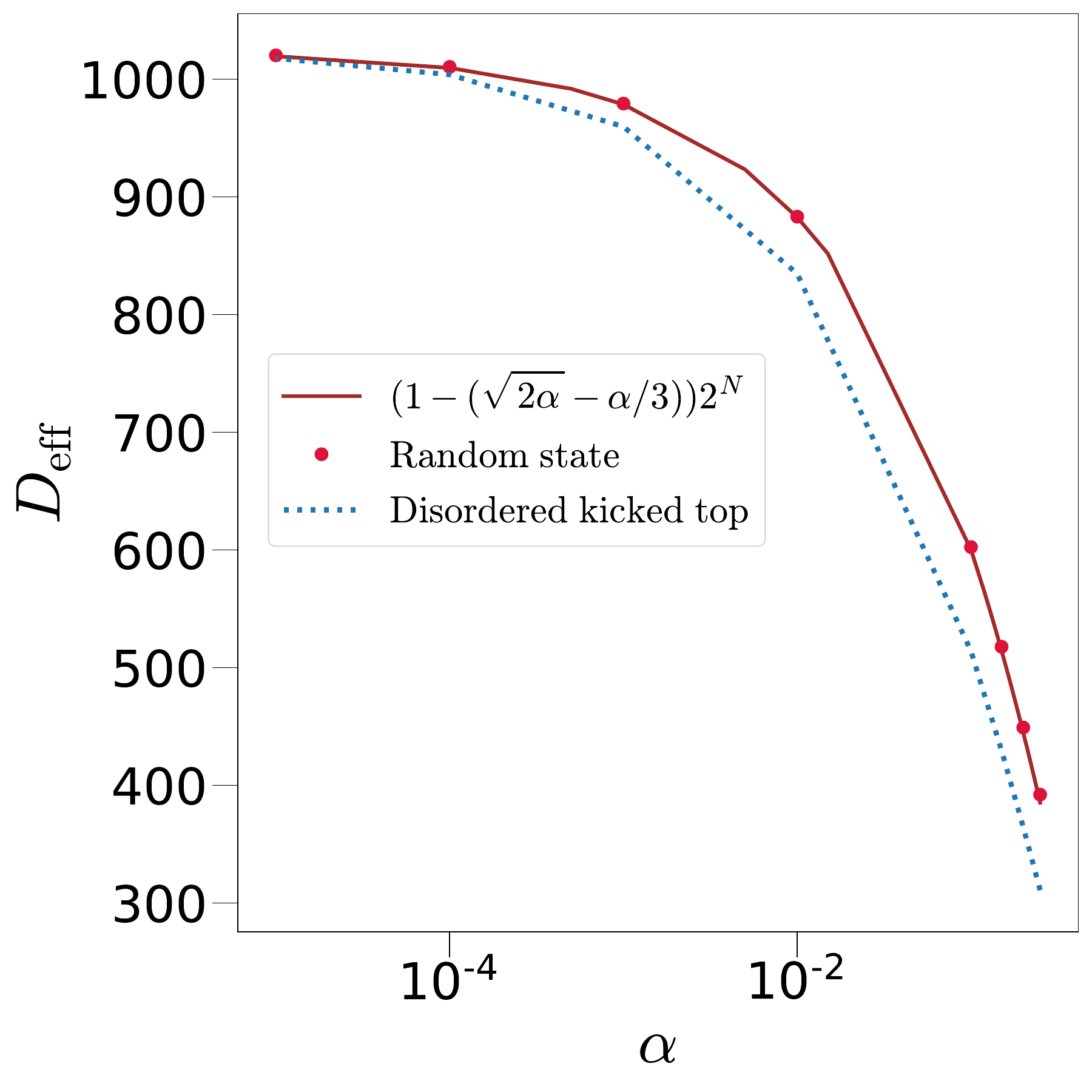}
    \caption{The dotted line shows the variation of $D_{\rm eff}$ with respect to $\alpha$ for $N=10$, $k=6$ and $w=5.0$ for disordered kicked top. The line is the analytical expression given by $(1-\sqrt{2\alpha}+\alpha/3)2^{N}$. The red dots (\textcolor{crimson}{$\bullet$}) on the line corresponds to the numerically obtained average random state values corresponding to the same Hilbert space dimensionality (1024). }
    \label{Deff_alpha}

\end{figure}

{\color{black}
\section{Random states $D_{\rm eff}$ from order statistics}
\label{appDeff}
Recall that in a $N$ dimensional Hilbert space, if the amplitude of the components of a normalized state were ordered in decreasing magnitude as $|c_1|^2\geq \cdots \geq |c_N|^2$, $D_{\rm eff}$ is the smallest $K$ such that $\sum_{i=1}^K |c_i|^2=1-\alpha$ for some fixed small $\alpha \ll 1$. We now wish to estimate how $K$ depends on $\alpha$ for Haar random states.

Consider a unit stick that is randomly broken into $N$ pieces. Let the broken sticks be arranged in the decreasing order of their length with $t_i$ as the length of the $i-$th largest piece so that
\begin{equation}
\sum_{i=1}^{N}t_{i}=1      \qquad 0 \leq t_i \leq 1 .
\end{equation}
The expectation value $\mathbb{E}(t_i)$ of $i-$th piece of the stick is given by \cite{David_orderstats}:
\begin{equation}
\mathbb{E}[t_i]=\frac{1}{N}\sum_{j=i}^{N}\frac{1}{j}.
\end{equation}

Thus,
\begin{align}
\mathbb{E}\left [\sum_{i=1}^{K} t_i \right] &= \frac{1}{N} \bigg( H_N + \left(H_N - 1\right) + \left(H_N - 1 - \frac{1}{2}\right) + \cdots \nonumber \\ 
&\quad + \left(H_N - 1 - \frac{1}{2} \cdots - \frac{1}{K-1} \right) \bigg), \nonumber \\
%&=\frac{1}{N}\left(KH_{N}-(K-1)-(K-2)\frac{1}{2}\right) \cdots \nonumber \\
%&\quad \left ((K-(K-1))\frac{1}{K-1} \right), \nonumber \\
&=\frac{1}{N} \left (KH_{N}-K(1+\frac{1}{2}+ \cdots \frac{1}{K})+K)\right), \nonumber \\
&=\frac{K}{N}\left(H_{N}-H_{K}  +1   \right).
\end{align}
Connecting this to the evaluation of $D_{\rm eff}$ for random states, which also involves the sum of first $K$ largest terms, i.e., $\sum_{i=1}^{K}|c_{i}|^{2}=1-\alpha$, we get:\\
\begin{align}
\label{eq_hnhm}
\mathbb{E}\left [\sum_{i=1}^{K} t_i \right] &= \frac{K}{N}\left(H_{N}-H_{K}  +1   \right) = 1-\alpha.
\end{align}

For a given $\alpha$ we need to find $K$.
Using $H_{N} - H_{K} \approx\ln N - \ln K$ and assuming a linear relation $K=\beta N$, Eq. \ref{eq_hnhm} can be rewritten as
\begin{align}
\ln \left (\frac{1}{\beta}\right)+1 &=\frac{1}{\beta}(1-\alpha),
\nonumber \\
\beta -\beta \ln \beta &= 1-\alpha.
\end{align}
Substituting $\beta=1-\delta$ and retaining terms up to $\delta^3$ in the expansion of $\ln (1-\delta)$ where $(\delta\ll1)$, we get $\delta^3+3\delta^2=6\alpha$. This can be solved by expanding $\delta=a_0 \sqrt{\alpha} +a_1 \alpha +O(\alpha^{3/2})$ and finding $a_0$ and $a_1$ by the usual matching of terms of the same order on both sides. We finally get  that
\begin{equation}
    \delta=\sqrt{2 \alpha} -\alpha/3 +O(\alpha^{3/2}).
\end{equation}
\begin{comment}
\begin{align}
%(1-\delta)-(1-\delta)\ln(1-\delta)=&1-\alpha, \nonumber\\
%(1-\delta)(1-(-\delta-\frac{\delta^2}{2}-\frac{\delta^3}{3}))=&1-\alpha ,\nonumber \\
\delta^3+3\delta^2=&6\alpha. 
\end{align}

Expressing this in depressed cubic form by the substitution $\delta=y-1$, gives
\begin{align}
y^3-3y+2=6\alpha . \nonumber \\
\end{align}
The real root of the above equation is given by:
\begin{align}
y&=\sqrt[3]{-1+3\alpha+\sqrt{-6\alpha+9\alpha^2}}+ \nonumber \\
& \qquad \sqrt[3]{-1+3\alpha-\sqrt{-6\alpha+9\alpha^2}} \nonumber\\
\end{align}
which in the limit of small $\alpha$ can be approximated as
\begin{align}
y &\approx 1+\sqrt{2\alpha}-\frac{\alpha}{3}+ \cdots
\end{align}
Since $\delta=y-1 $, we get
\begin{align}
\delta\approx\sqrt{2\alpha}-\frac{\alpha}{3}. \nonumber \\
\end{align}
\end{comment}Thus, the number of components required to reach a normalization of $1-\alpha$ is given by
$K=N(1-\sqrt{2\alpha}+\alpha/3+O(\alpha^{3/2}))$

Hence, $D_{\rm{eff}}$ for random states of $N$ qubits, denoted as $\overline{D_{\rm eff}}$, is approximately given by
$(1-\sqrt{2\alpha}+\alpha/3)2^{N}$, as the dimensionality of the Hilbert space is $2^N$. The variation of $D_{\rm eff}$ with $\alpha$ for a large value of the kicked-top parameter $k$ and large disorder limit is shown in Fig. \ref{Deff_alpha}. While following the derived analytical expression, there is a consistent deviation from random states and indicates a small measure of localization. Measures based on order statistics, especially extreme value statistics, have been found to be highly sensitive to deviations from random states, resulting in subtle discrepancies when the states are not ``perfectly" random as in these finite systems. Related studies focusing on extreme value statistics based measures, which reveal such deviations are discussed in Ref. \cite{ent_coupled_rotors, coupled_kicked_top}. }

%\newpage
%\bibliography{ref}
%\nocite{*}
%%%%%%%%%%%%%%%%%%%%%%%%%%%%%%%%%%%%%%%%%%%%%%%%%%%%%%%%%%%%%%%
%\vspace{-\baselineskip}
%\begin{thebibliography}{99}
%\vspace{-\baselineskip}
%%%%%%%%%%%%%%%%%%%%%%%%%%%%%%%%%%%%%%%%%%%%%%%%%%%%%%%%%%%%%%%

%merlin.mbs apsrev4-1.bst 2010-07-25 4.21a (PWD, AO, DPC) hacked
%Control: key (0)
%Control: author (8) initials jnrlst
%Control: editor formatted (1) identically to author
%Control: production of article title (-1) disabled
%Control: page (0) single
%Control: year (1) truncated
%Control: production of eprint (0) enabled
%

\end{document}